\documentclass[12pt,nofootinbib]{revtex4}

\usepackage{amsmath,amssymb}
\usepackage{graphicx}

\usepackage{setspace}
\usepackage{color}

\begin{document}

\title{Higgs and $W$ boson spectrum from lattice simulations}

\author{Mark Wurtz and Randy Lewis}
\affiliation{Department of Physics and Astronomy, York University, Toronto,
             Ontario, M3J 1P3, Canada}

\begin{abstract}
The spectrum of energy levels is computed for all available angular momentum and parity quantum numbers in the SU(2)-Higgs model, with parameters chosen to match experimental data from the Higgs-$W$ boson sector of the standard model. Several multiboson states are observed, with and without linear momentum, and all are consistent with weakly interacting Higgs and $W$ bosons. The creation operators used in this study are gauge-invariant so, for example, the Higgs operator is quadratic rather than linear in the Lagrangian's scalar field.
\end{abstract}

\maketitle

\section{Introduction}

The complex scalar doublet of the standard model accommodates all of the 
necessary masses for elementary particles.  A testable prediction of this
theory is the presence of a fundamental scalar particle: the Higgs boson.
Recently, ATLAS and CMS have discovered a Higgs-like boson with a mass near 125
GeV \cite{Aad:2012tfa,Chatrchyan:2012ufa}.

Lattice simulations of the scalar doublet with the SU(2) gauge part of the
electroweak theory give a nonperturbative description of the Higgs mechanism.
Early studies \cite{Fradkin:1978dv,Osterwalder:1977pc,Lang:1981qg,Seiler:1982pw,
Kuhnelt:1983mw,Montvay:1984wy,Jersak:1985nf,Evertz:1985fc,Gerdt:1984ft,
Langguth:1985dr,Montvay:1985nk,Langguth:1987vf,Evertz:1989hb,Hasenfratz:1987uc}
revealed two regions in the phase diagram:
the Higgs region with three massive vector
bosons and a single Higgs particle, and the confinement region with QCD-like
bound states of the fundamental fields.  These two regions are partially
separated by a first-order phase transition, but are analytically connected
beyond the phase transition's end point.
Subsequent lattice studies of the SU(2)-Higgs model have explored the 
electroweak finite-temperature phase
transition \cite{Jansen:1995yg,Rummukainen:1996sx,Laine:1998jb,Fodor:1999at}
and recent work has incorporated additional scalar
doublets \cite{Wurtz:2009gf,Lewis:2010ps}.

In the present work, we calculate the spectrum of the standard SU(2)-Higgs
model at zero temperature in the Higgs region of the phase diagram.
As already mentioned, there will be a Higgs boson ($H$) and three
massive vector bosons ($W^1$, $W^2$ and $W^3$), but the spectrum contains
much more than this.

For comparison recall the well-known case of QCD, which has a
small number of fields in the Lagrangian (gluons and quarks) and a huge
number of particles in the spectrum (glueballs and hadrons).
The glueballs and hadrons are created by gauge-invariant operators but the
gluons and quarks correspond to gauge-dependent fields in the Lagrangian.
The spectrum of the SU(2)-Higgs model is similar, at least in the confinement
region: the Lagrangian contains gauge fields and a doublet of scalar fields,
but lattice simulations suggest a dense spectrum of ``W-balls'' and
``hadrons.''  (For lattice studies of the spectrum in 2+1 dimensions, see
Refs.~\cite{Philipsen:1996af,Philipsen:1997rq}.)

It is interesting to consider the spectrum in the Higgs region of the
phase diagram.
At weak coupling (which is directly relevant to the actual experimental
situation), one might anticipate one Higgs boson, three vector bosons, and
nothing else.  On the other hand, since the Higgs region and the confinement
region are truly a single phase, one might wonder whether
the rich spectrum of
confinement-region states will persist into the Higgs region, though smoothly
rearranged in some way.
An appealing view can be found in Refs.~\cite{Philipsen:1996af,Philipsen:1997rq}
where the smooth transition from confinement region to Higgs region was
observed for an SU(2)-Higgs model in 2+1 dimensions.
Reference~\cite{Philipsen:1997rq} describes the results in terms of a flux loop
that is completely stable in the pure gauge theory but can decay in the
confinement region of the SU(2)-Higgs phase diagram.
When approaching the analytic pathway into the Higgs region,
such decays become so
rapid that the particle description loses its relevance, leaving the
Higgs region with the simple spectrum of Higgs and $W$ bosons.
Reference~\cite{Philipsen:1997rq} concludes by emphasizing the usefulness of a future
study of multiparticle states in the Higgs region.

In practice, even a simple spectrum of four bosons ($W^1$, $W^2$, $W^3$, $H$)
will be accompanied by a tower of multiparticle states
($WW$, $WH$, $HH$, $WWW$, \dots) that is consistent with conservation of
weak isospin, angular momentum and parity.
Therefore a thorough lattice study of the spectrum will always involve many
states appearing with many different quantum numbers.
In general, these could be bound states and/or scattering states, and
there is a history of nonlattice attempts to determine whether
a pair of Higgs bosons might form a bound state
\cite{Cahn:1983vi,Contogouris:1988rd,Grifols:1991gw,Rupp:1991bb,DiLeo:1994xc,Clua:1995ni,DiLeo:1996aw,Siringo:2000kh}.

The existence of nonperturbative states for
$\phi^4$ theory in 2+1 dimensions has support from lattice simulations
\cite{Caselle:2000yx,Caselle:1999tm}.
Attempts for the 3+1 dimensional SU(2)-Higgs model \cite{Maas:2012tj,Maas:2012zf}
(see for example Fig.~3 of Ref.~\cite{Maas:2012zf}) indicate that
the task of computing the Higgs-region spectrum with sufficient precision 
to observe and identify more than the most basic states is a significant challenge.
We have had success in this endeavor, which is the theme of the present work.

Section~\ref{sec:simulation} describes the method used to create the lattice
ensembles.  Section~\ref{sec:operators} develops a set of creation operators
that is able to probe all quantum numbers $I(\Lambda^P)$, where $I$
denotes weak isospin, $P$ is parity, and $\Lambda$
is a lattice representation corresponding to angular momentum.
Section~\ref{sec:analysis} explains how the variational method was used for
analysis of the lattice data.
Section~\ref{sec:spectrum} presents the energy spectrum that was obtained
from our lattice simulations.
Section~\ref{sec:biggerlattice} examines the effects on the spectrum
of increasing the lattice volume.
Section~\ref{sec:infiniteHiggs} reports on a simulation with a much larger
Higgs mass, so that changes in the energy spectrum can be observed and understood.
Section~\ref{Two-Particle Operators} describes the construction of two-particle
operators and uses them to extend the observed energy spectrum.
Concluding remarks are contained in Sec.~\ref{sec:conclusions}.

\section{Simulation Details}\label{sec:simulation}

The discretized SU(2)-Higgs action used for lattice simulations is given by
\begin{eqnarray}
S[U,\phi] &=& \sum_x \left\{ \beta \sum_{\mu<\nu} \left[ 1 -  \tfrac{1}{2} \operatorname{Tr} \left( U_\mu(x) U_\nu(x+\hat\mu)U_\mu^\dag(x+\hat\nu)U_\nu^\dag(x) \right)  \right]  \right.  \nonumber \\
&& + \tfrac{1}{2} \operatorname{Tr} \left(\phi^\dag(x)\phi(x)\right) + \lambda \left( \tfrac{1}{2} \operatorname{Tr} \left(\phi^\dag(x)\phi(x)\right) - 1 \right)^2 \nonumber \\
&& \left. - \kappa \sum_{\mu=1}^4 \operatorname{Tr} \left( \phi^\dag(x)U_\mu(x) \phi(x+\hat\mu) \right) \right\}  \,\, ,
\end{eqnarray}
where $U_\mu(x) = e^{iag_0A_\mu(x)}$ is the gauge field, $\phi(x)$ is the scalar field in $2\times2$ matrix form, $\beta = \frac{4}{g_0^2}$ is the gauge coupling, $\kappa = \frac{1-2\lambda}{8+a^2\mu_0^2}$ is the hopping parameter (related to the inverse bare mass squared), and $\lambda = \kappa^2\lambda_0$ is the scalar self-coupling.  The $2\times2$ complex scalar field contains only four degrees of freedom because of a relation involving a Pauli matrix,
\begin{align}
\sigma_2\phi(x)\sigma_2 = \phi^*(x) \,,
\end{align}
and is written as $\phi(x) = \rho(x)\alpha(x)$, where $\rho(x)>0$ is called the scalar length and $\alpha(x)\in SU(2)$ is the scalar's angular component.
We refer to $\phi(x)$ as the scalar field rather than the Higgs field,
reserving the ``Higgs'' label for the physical particle which, as discussed
in Sec.~\ref{sec:operators}, is quadratic in the scalar field.

Our simulations are performed in the Higgs region of the phase diagram, with a gauge coupling near the physical value $g_0^2\approx\frac{4\pi\alpha}{\sin^2\theta_W}\approx\frac{4\pi\alpha}{1-m_W^2/m_Z^2}\approx 0.5$, corresponding to $\beta=8$, which is in the weak coupling region.  The remaining parameters are tuned to $\kappa=0.131$ and $\lambda=0.0033$ to give a Higgs mass near the physical value of $\sim$125 GeV and a reasonable lattice spacing.  The number of lattice sites is $20^3\times40$ (where the longer direction is Euclidean time) and $24^3\times48$, and the scale is set with the $W$ mass defined to be 80.4 GeV.
For comparison, separate simulations are carried out with $\kappa=0.4$ and
$\lambda=\infty$.

Although $\phi^4$ theories are trivial, the standard model can be viewed as  
an effective field theory up to some finite cutoff.  The calculations  
presented in this paper are at a cutoff of approximately $1/a=400$ GeV.  
Even though the continuum limit is problematic in a trivial theory,  
simulations at an appropriately-large cutoff are sufficient to produce  
phenomenological results.

Standard heatbath and over-relaxation algorithms \cite{Creutz:1980zw,Creutz:1984mg,Kennedy:1985nu,Creutz:1987xi,Bunk:1994xs,Fodor:1994sj} were used for the Monte Carlo update of the gauge and scalar fields.
Define one sweep to mean an update at all sites across the lattice.
Then our basic update step is one gauge heatbath sweep
followed by two scalar heatbath sweeps
followed by one gauge over-relaxation sweep
followed by four scalar over-relaxation sweeps.
Ten of these basic update steps are performed between the calculation of
lattice observables.  Any remaining autocorrelation is handled by binning the
observable.

Stout link smearing \cite{Morningstar:2003gk} and scalar smearing \cite{Bulava:2009jb,Peardon:2009gh} are used to improve the lattice operators, reduce statistical fluctuations, and construct a large basis of operators.  For the gauge links, one stout-link iteration is given by
\begin{align}
U^{(n+1)}_\mu(x) &= \exp\left\{-r_{\rm stout} \, Q^{(n)}_\mu(x)\right\} U^{(n)}_\mu(x)  \quad , \quad \mu \neq 4 \\
Q^{(n)}_\mu(x) &= \frac{1}{2} \sum_{\nu \neq \mu, \nu \neq 4} \left\{ U^{(n)}_\mu(x) U^{(n)}_\nu(x+\hat{\mu}) U^{(n)\dag}_\mu(x+\nu) U^{(n)\dag}_\nu(x) \right. \nonumber \\
& \left. \qquad\qquad + U^{(n)}_\mu(x) U^{(n)\dag}_\nu(x+\hat{\mu}-\hat{\nu}) U^{(n)\dag}_\mu(x-\hat{\nu}) U^{(n)}_\nu(x-\hat{\nu}) \right\} - \text{h.c.}
\end{align}
where $r_{\rm stout}$ is the stout link smearing parameter.  Only the spatial links are smeared, and only in the spatial direction.  The final stout links $\tilde{U}$ are given after a number of successive smearing iterations
\begin{align}
U = U^{(0)} \rightarrow U^{(1)} \rightarrow U^{(2)} \rightarrow \cdots \rightarrow U^{(n_{\rm stout})} = \tilde{U} \,\, .
\end{align}
The smearing for the scalar field uses the lattice Laplacian $\Delta$,
\begin{align}
\phi^{(n+1)}(x) &= \left(1 + r_{\rm smear}\Delta\right)\phi^{(n)}(x) \\
&= \phi^{(n)}(x) + r_{\rm smear}\sum_{\mu=1}^3\left\{ \tilde{U}_\mu(x)\phi^{(n)}(x+\hat{\mu}) - 2\phi^{(n)}(x) + \tilde{U}^\dag_\mu(x-\hat{\mu})\phi^{(n)}(x-\hat{\mu}) \right\} \,\, ,
\end{align}
where $r_{\rm smear}$ is the scalar smearing parameter.  Note that the stout links $\tilde{U}$ are used for scalar smearing, and only in spatial directions.  The final smeared scalar fields $\tilde{\phi}$ are given by
\begin{align}
\phi = \phi^{(0)} \rightarrow \phi^{(1)} \rightarrow \phi^{(2)} \rightarrow \cdots \rightarrow \phi^{(n_{\rm smear})} = \tilde{\phi} \,\, .
\end{align}

\section{Primary operators}\label{sec:operators}

This study begins with two basic options for gauge-invariant operators, the first being two scalar fields connected by a string of gauge links, and the second being a closed loop of gauge links.  Use of stout links and smeared scalar fields within those operators enables many different possible gauge link paths and scalar field separations to be included.  To obtain information about continuum angular momentum from a lattice simulation, there is a well-known correspondence with irreducible representations (irreps) of the octahedral group of rotations \cite{Johnson:1982yq,Berg:1982kp}, as shown in Table~\ref{irrep_spin_table}.

\begin{table}[b]
\caption{The number of copies of each irreducible representation $\Lambda$
         for each continuum spin $J$.}
\label{irrep_spin_table}
\begin{tabular}{l|cccccccc} 
$\Lambda$ & \multicolumn{8}{c}{$J$} \\
\cline{2-9}
 & 0 & 1 & 2 & 3 & 4 & 5 & 6 & $\dots$ \\
\hline 
$A_1$ & 1 & 0 & 0 & 0 & 1 & 0 & 1 & $\dots$ \\
$A_2$ & 0 & 0 & 0 & 1 & 0 & 0 & 1 & $\dots$ \\
$E$   & 0 & 0 & 1 & 0 & 1 & 1 & 1 & $\dots$ \\
$T_1$ & 0 & 1 & 0 & 1 & 1 & 2 & 1 & $\dots$ \\
$T_2$ & 0 & 0 & 1 & 1 & 1 & 1 & 2 & $\dots$
\end{tabular} 
\end{table}

The simplest gauge-invariant operator that can be constructed from scalar fields is the Higgs length operator
\begin{align}
H(t) =  \frac{1}{2} \operatorname{Tr}\sum_{\vec{x}}\phi^\dag(x)\phi(x) = \sum_{\vec{x}} \rho^2(x) \,\, ,
\end{align}
where the sum includes all spatial sites at a single Euclidean time.
The $H(t)$ operator transforms according to the $\Lambda^P=A_1^+$ irrep and thus couples to the spin-0 Higgs state.
Notice that the Higgs operator is quadratic in the scalar field $\phi(x)$,
as is familiar from the earliest SU(2)-Higgs model lattice simulations
\cite{Fradkin:1978dv,Osterwalder:1977pc,Lang:1981qg,Seiler:1982pw,
Kuhnelt:1983mw,Montvay:1984wy,Jersak:1985nf,Evertz:1985fc,Gerdt:1984ft,
Langguth:1985dr,Montvay:1985nk,Langguth:1987vf,Evertz:1989hb,Hasenfratz:1987uc}.

The simplest operator that couples to the $W$ particle is the isovector gauge-invariant link
\begin{align}
W^a_\mu(t) = \frac{1}{2} \operatorname{Tr} \sum_{\vec{x}} -i\sigma^a \phi^\dag(x) U_\mu(x) \phi(x+\hat{\mu}) \,\, ,
\end {align}
which belongs to the $\Lambda^P=T_1^-$ irrep.  Notice that, in general, an isovector operator does not have definite charge conjugation.  The operator $W^a_\mu(t)$, for example, transforms under charge conjugation as $(W^1_\mu,W^2_\mu,W^3_\mu) \rightarrow (-W^1_\mu,+W^2_\mu,-W^3_\mu)$.  Clearly, if the operator $W_\mu^a$ is given an arbitrary isospin rotation it will not be an eigenfunction of charge conjugation.  Therefore charge conjugation is not helpful for the present work.

\begin{figure}[tb]
\includegraphics[scale=1.0]{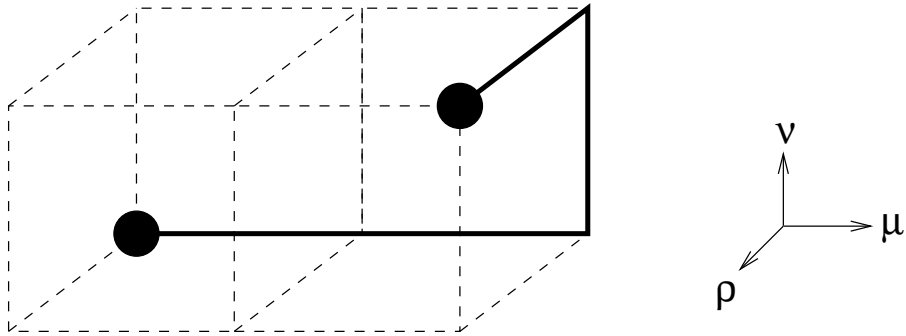}
\caption{Sketch of the two-scalar-field operator $L_{\mu\nu\rho}$.  The two dots at the ends of $L_{\mu\nu\rho}$ represent the scalar fields.}
\label{figure_link}
\end{figure}

Other irreps can be obtained by considering more complicated operators.  The gauge-invariant link operator
\begin{align}
L^\phi_{\mu\nu\rho}(t) = \sum_{\vec{x}} \phi^\dag(x) U_\mu(x) U_\mu(x+\hat{\mu}) U_\nu(x+2\hat{\mu}) U_\rho(x+2\hat{\mu}+\hat{\nu}) \phi(x+2\hat{\mu}+\hat{\nu}+\hat{\rho}) \,,  \label{asymtwistedlinkphi}
\end{align}
shown in Fig.~\ref{figure_link},
has 48 possible orientations and is one of the simplest two-scalar-field operators that couples to all of the $I(\Lambda^P)$ channels.  Also considered is the gauge-invariant link constructed using SU(2)-``angular'' components of the scalar field
\begin{align}
L^\alpha_{\mu\nu\rho}(t) = \sum_{\vec{x}} \alpha^\dag(x) U_\mu(x) U_\mu(x+\hat{\mu}) U_\nu(x+2\hat{\mu}) U_\rho(x+2\hat{\mu}+\hat{\nu}) \alpha(x+2\hat{\mu}+\hat{\nu}+\hat{\rho}) \,\, ,  \label{asymtwistedlinkalpha}
\end{align}
which has exactly the same rotational symmetries as $L^\phi_{\mu\nu\rho}(t)$.  Useful linear combinations of $L_{\mu\nu\rho}(t)$ (dropping the $\phi$, $\alpha$ and $t$ symbols for brevity) are given by
\begin{align}
A^+_{\mu\nu\rho} &= L_{+\mu+\nu+\rho} + L_{+\mu+\nu-\rho} + L_{+\mu-\nu+\rho} + L_{+\mu-\nu-\rho} \nonumber \\
                 &+ L_{-\mu+\nu+\rho} + L_{-\mu+\nu-\rho} + L_{-\mu-\nu+\rho} + L_{-\mu-\nu-\rho} \label{operatorA+} \\
A^-_{\mu\nu\rho} &= L_{+\mu+\nu+\rho} - L_{+\mu+\nu-\rho} - L_{+\mu-\nu+\rho} + L_{+\mu-\nu-\rho} \nonumber \\
                 &- L_{-\mu+\nu+\rho} + L_{-\mu+\nu-\rho} + L_{-\mu-\nu+\rho} - L_{-\mu-\nu-\rho} \label{operatorA-} \\
B^+_{\mu\nu\rho} &= L_{+\mu+\nu+\rho} - L_{+\mu+\nu-\rho} - L_{+\mu-\nu+\rho} + L_{+\mu-\nu-\rho} \nonumber \\
                 &+ L_{-\mu+\nu+\rho} - L_{-\mu+\nu-\rho} - L_{-\mu-\nu+\rho} + L_{-\mu-\nu-\rho} \label{operatorB+} \\
B^-_{\mu\nu\rho} &= L_{+\mu+\nu+\rho} + L_{+\mu+\nu-\rho} + L_{+\mu-\nu+\rho} + L_{+\mu-\nu-\rho} \nonumber \\
                 &- L_{-\mu+\nu+\rho} - L_{-\mu+\nu-\rho} - L_{-\mu-\nu+\rho} - L_{-\mu-\nu-\rho} \label{operatorB-} \\
C^+_{\mu\nu\rho} &= L_{+\mu+\nu+\rho} + L_{+\mu+\nu-\rho} - L_{+\mu-\nu+\rho} - L_{+\mu-\nu-\rho} \nonumber \\
                 &- L_{-\mu+\nu+\rho} - L_{-\mu+\nu-\rho} + L_{-\mu-\nu+\rho} + L_{-\mu-\nu-\rho} \label{operatorC+} \\
C^-_{\mu\nu\rho} &= L_{+\mu+\nu+\rho} + L_{+\mu+\nu-\rho} - L_{+\mu-\nu+\rho} - L_{+\mu-\nu-\rho} \nonumber \\
                 &+ L_{-\mu+\nu+\rho} + L_{-\mu+\nu-\rho} - L_{-\mu-\nu+\rho} - L_{-\mu-\nu-\rho} \label{operatorC-} \\
D^+_{\mu\nu\rho} &= L_{+\mu+\nu+\rho} - L_{+\mu+\nu-\rho} + L_{+\mu-\nu+\rho} - L_{+\mu-\nu-\rho} \nonumber \\
                 &- L_{-\mu+\nu+\rho} + L_{-\mu+\nu-\rho} - L_{-\mu-\nu+\rho} + L_{-\mu-\nu-\rho} \label{operatorD+} \\
D^-_{\mu\nu\rho} &= L_{+\mu+\nu+\rho} - L_{+\mu+\nu-\rho} + L_{+\mu-\nu+\rho} - L_{+\mu-\nu-\rho} \nonumber \\
                 &+ L_{-\mu+\nu+\rho} - L_{-\mu+\nu-\rho} + L_{-\mu-\nu+\rho} - L_{-\mu-\nu-\rho} \label{operatorD-}
\end{align}
and Table~\ref{operator_table} shows how to construct operators of any irrep and parity.  Note that operators $A^+_{\mu\nu\rho}$, $B^+_{\mu\nu\rho}$, $C^+_{\mu\nu\rho}$ and $D^+_{\mu\nu\rho}$ are even under parity, whereas $A^-_{\mu\nu\rho}$, $B^-_{\mu\nu\rho}$, $C^-_{\mu\nu\rho}$ and $D^-_{\mu\nu\rho}$ are odd.  The operators $A^\pm_{\mu\nu\rho}$ belong to the $A_1$, $A_2$ and $E$ irreps, whereas $B^\pm_{\mu\nu\rho}$, $C^\pm_{\mu\nu\rho}$ and $D^\pm_{\mu\nu\rho}$ belong to the $T_1$ and $T_2$ irreps.

\begin{table}[tb]
\caption{Linear combinations of operators that give any irrep and parity.  The multiplicity, mult$(\Lambda^P)$, is shown for each case.}
\label{operator_table}
\begin{tabular}{l|c|l} 
$\Lambda^P$ & mult$(\Lambda^P)$ & \multicolumn{1}{c}{operators} \\
\hline
$A_1^{+}$ & 1 & $A^+_{123} + A^+_{231} + A^+_{312} + A^+_{132} + A^+_{213} + A^+_{321}$ \\
$A_1^{-}$ & 1 & $A^-_{123} + A^-_{231} + A^-_{312} - A^-_{132} - A^-_{213} - A^-_{321}$ \\
$A_2^{+}$ & 1 & $A^+_{123} + A^+_{231} + A^+_{312} - A^+_{132} - A^+_{213} - A^+_{321}$ \\
$A_2^{-}$ & 1 & $A^-_{123} + A^-_{231} + A^-_{312} + A^-_{132} + A^-_{213} + A^-_{321}$ \\
$E^{+}$   & 2 & $\left\{(A^+_{123} - A^+_{231} + A^+_{132} - A^+_{213}) / \sqrt{2}, \right.$ \\
          &   & $\left. (A^+_{123} + A^+_{231} -2A^+_{312} + A^+_{132} + A^+_{213} -2A^+_{321}) / \sqrt{6} \right\}$ \\
          &   & $\left\{(A^+_{123} - A^+_{231} - A^+_{132} + A^+_{213}) / \sqrt{2}, \right.$ \\
          &   & $\left. (A^+_{123} + A^+_{231} -2A^+_{312} - A^+_{132} - A^+_{213} +2A^+_{321}) / \sqrt{6} \right\}$ \\
$E^{-}$   & 2 & $\left\{(A^-_{123} - A^-_{231} + A^-_{132} - A^-_{213}) / \sqrt{2}, \right.$ \\
          &   & $\left. (A^-_{123} + A^-_{231} -2A^-_{312} + A^-_{132} + A^-_{213} -2A^-_{321}) / \sqrt{6} \right\}$ \\
          &   & $\left\{(A^-_{123} - A^-_{231} - A^-_{132} + A^-_{213}) / \sqrt{2}, \right.$ \\
          &   & $\left. (A^-_{123} + A^-_{231} -2A^-_{312} - A^-_{132} - A^-_{213} +2A^-_{321}) / \sqrt{6} \right\}$ \\
$T_1^{+}$ & 3 & $\left\{ B^+_{123} - B^+_{132} \,,\, B^+_{231} - B^+_{213} \,,\, B^+_{312} - B^+_{321} \right\}$ \\
          &   & $\left\{ C^+_{123} - C^+_{213} \,,\, C^+_{231} - C^+_{321} \,,\, C^+_{312} - C^+_{132} \right\}$ \\
          &   & $\left\{ D^+_{123} - D^+_{321} \,,\, D^+_{231} - D^+_{132} \,,\, D^+_{312} - D^+_{213} \right\}$ \\
$T_1^{-}$ & 3 & $\left\{ B^-_{123} + B^-_{132} \,,\, B^-_{231} + B^-_{213} \,,\, B^-_{312} + B^-_{321} \right\}$ \\
          &   & $\left\{ C^-_{123} + C^-_{321} \,,\, C^-_{231} + C^-_{132} \,,\, C^-_{312} + C^-_{213} \right\}$ \\
          &   & $\left\{ D^-_{123} + D^-_{213} \,,\, D^-_{231} + D^-_{321} \,,\, D^-_{312} + D^-_{132} \right\}$ \\
$T_2^{+}$ & 3 & $\left\{ B^+_{123} + B^+_{132} \,,\, B^+_{231} + B^+_{213} \,,\, B^+_{312} + B^+_{321} \right\}$ \\
          &   & $\left\{ C^+_{123} + C^+_{213} \,,\, C^+_{231} + C^+_{321} \,,\, C^+_{312} + C^+_{132} \right\}$ \\
          &   & $\left\{ D^+_{123} + D^+_{321} \,,\, D^+_{231} + D^+_{132} \,,\, D^+_{312} + D^+_{213} \right\}$ \\
$T_2^{-}$ & 3 & $\left\{ B^-_{123} - B^-_{132} \,,\, B^-_{231} - B^-_{213} \,,\, B^-_{312} - B^-_{321} \right\}$ \\
          &   & $\left\{ C^-_{123} - C^-_{321} \,,\, C^-_{231} - C^-_{132} \,,\, C^-_{312} - C^-_{213} \right\}$ \\
          &   & $\left\{ D^-_{123} - D^-_{213} \,,\, D^-_{231} - D^-_{321} \,,\, D^-_{312} - D^-_{132} \right\}$ \\
\end{tabular} 
\end{table}

\begin{figure}[tb]
\includegraphics[scale=1.0]{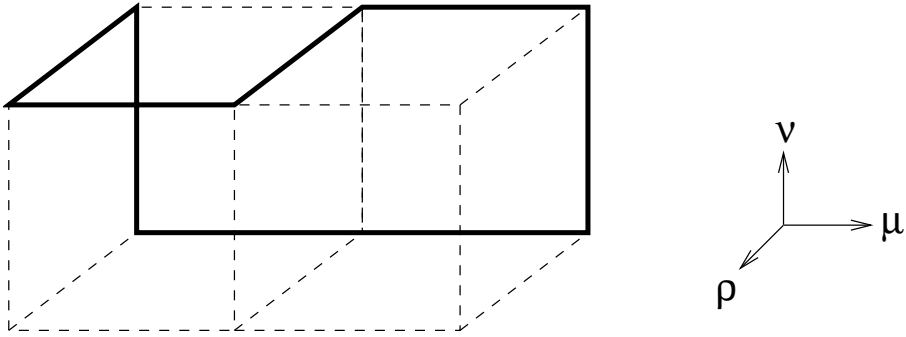}
\caption{The Wilson loop operator $W_{\mu\nu\rho}$ of Eq.~(\ref{wilsonloop}).}
\label{figure_wilson}
\end{figure}

\begin{figure}[tb]
\includegraphics[scale=1.0]{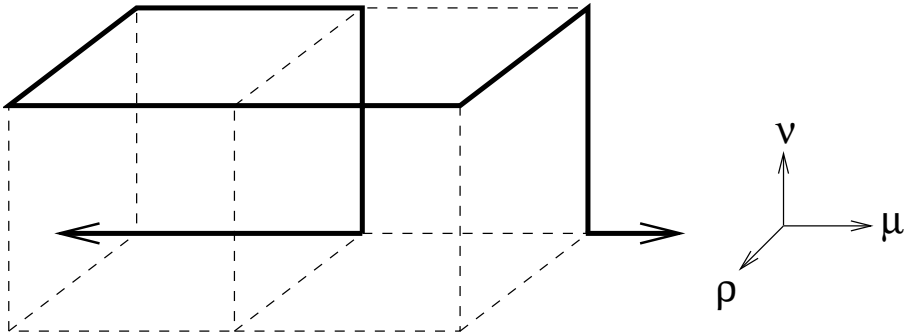}
\caption{The ``kinked'' Polyakov loop operator $P_{\mu\nu\rho}$ of Eq.~(\ref{polyakovloop}).}
\label{figure_polyakov}
\end{figure}

The operator $L_{\mu\nu\rho}$ consists of four gauge-invariant real components:
one is an isoscalar,
\begin{equation}
\frac{1}{2}{\rm Tr}(L_{\mu\nu\rho}) \,,
\end{equation}
and the other three form an isovector,
\begin{equation}
\frac{1}{2}{\rm Tr}(-i\sigma^aL_{\mu\nu\rho}) \,.
\end{equation}
In addition to the gauge-invariant link, which contains two scalar fields, there are operators that contain only gauge fields.  A Wilson loop is a gauge-invariant operator in which the path of gauge links returns to itself to form a closed loop.  A particular Wilson loop that couples to all available irreps is shown in Fig.~\ref{figure_wilson}.
Mathematically, it is
\begin{align}
W_{\mu\nu\rho}(t) = \frac{1}{2} \operatorname{Tr} & \sum_{\vec{x}} U_\mu(x) U_\mu(x+\hat{\mu}) U_\nu(x+2\hat{\mu}) U^\dag_\mu(x+\hat{\mu}+\hat{\nu}) \nonumber \\
& \quad \times U_\rho(x+\hat{\mu}+\hat{\nu}) U^\dag_\mu(x+\hat{\nu}+\hat{\rho}) U^\dag_\rho(x+\hat{\nu}) U^\dag_\nu(x) \label{wilsonloop}
\end{align}
which is operator \#4 in Table 3.2 of Ref.~\cite{Berg:1982kp} and has 48 different orientations.  A Polyakov loop is also a gauge-invariant closed loop, but it wraps around a boundary of the periodic lattice.  All irreps can be obtained from a Polyakov loop that contains a ``kink,'' denoted by $K_{\mu\nu\rho}$, such as the one shown in Fig.~\ref{figure_polyakov} which is
\begin{align}
P_{\mu\nu\rho}(t) &= \frac{1}{2} \operatorname{Tr} \sum_{\vec{x}} \left\{ \prod_{y_\mu < x_\mu} U_{\mu}(x+(y_\mu-x_\mu)\hat{\mu}) \right\}K_{\mu\nu\rho}(x)
\left\{ \prod_{y_\mu > x_\mu} U_{\mu}(x+(y_\mu-x_\mu)\hat{\mu}) \right\} \,, \label{polyakovloop} \\
K_{\mu\nu\rho}(x) &= U_\nu(x) U^\dag_\mu(x+\hat{\nu}-\hat{\mu}) U_\rho(x+\hat{\nu}-\hat{\mu}) U_\mu(x+\hat{\nu}-\hat{\mu}+\hat{\rho}) \nonumber \\
&\times U_\mu(x+\hat{\nu}+\hat{\rho}) U^\dag_\rho(x+\hat{\nu}+\hat{\mu}) U^\dag_\nu(x+\hat{\mu}) \,,
\end{align}
and has 48 different orientations.  The kink $K_{\mu\nu\rho}$ is inserted to fill the gap between points $x$ and $x+\hat{\mu}$ of an otherwise normal Polyakov loop.  All possible irreps and parities for $W_{\mu\nu\rho}$ and $P_{\mu\nu\rho}$ can be obtained from Table~\ref{operator_table} simply by replacing $L_{\mu\nu\rho}$ with $W_{\mu\nu\rho}$ or $P_{\mu\nu\rho}$ in Eqs.~\eqref{operatorA+} to \eqref{operatorD-}.  Since a Pauli matrix cannot be inserted into the trace of a closed loop operator made entirely of gauge links without destroying gauge invariance, there are no isovector Wilson or Polyakov loop operators.

\begin{figure}[tb]
\includegraphics[scale=0.6,clip=true]{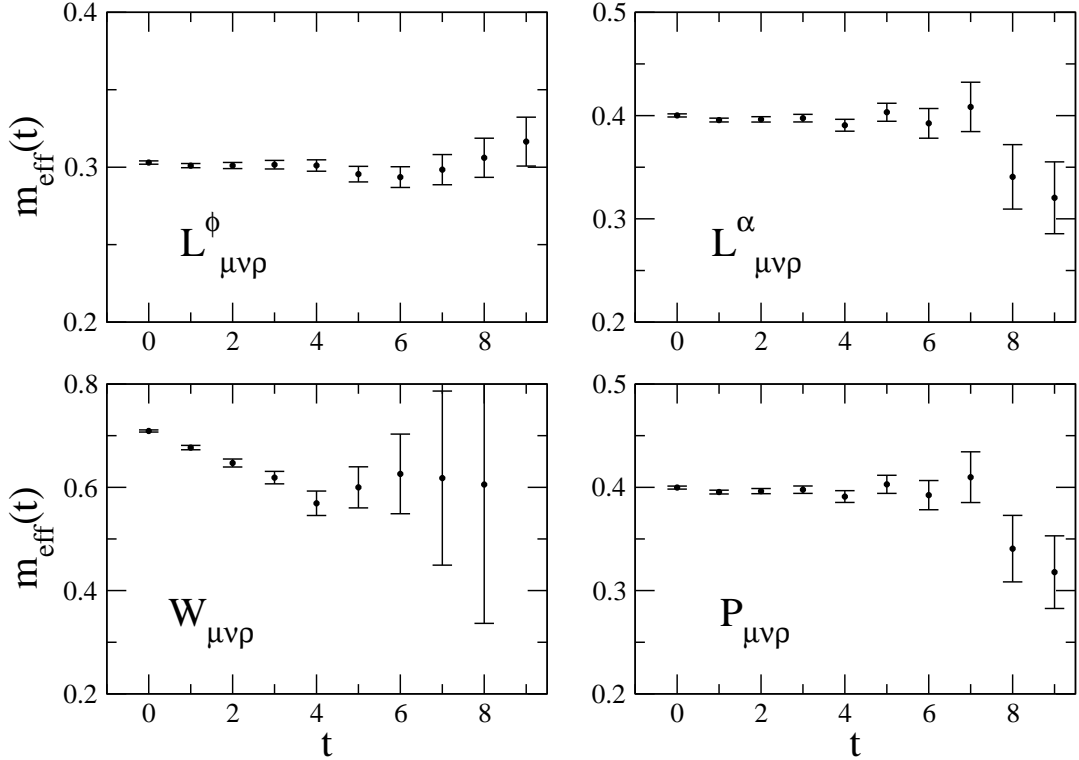}
\caption{Effective masses of the $I(\Lambda^P)=0(A_1^+)$ gauge-invariant link operators $L^\phi_{\mu\nu\rho}$ and $L^\alpha_{\mu\nu\rho}$, Wilson loop $W_{\mu\nu\rho}$ and Polyakov loop $P_{\mu\nu\rho}$ operators on a $20^3\times 40$ lattice with $\beta=8$, $\kappa=0.131$ and $\lambda=0.0033$.}
\label{graph_effmass_0A1+}
\end{figure}
\begin{figure}[tb]
\includegraphics[scale=0.6,clip=true]{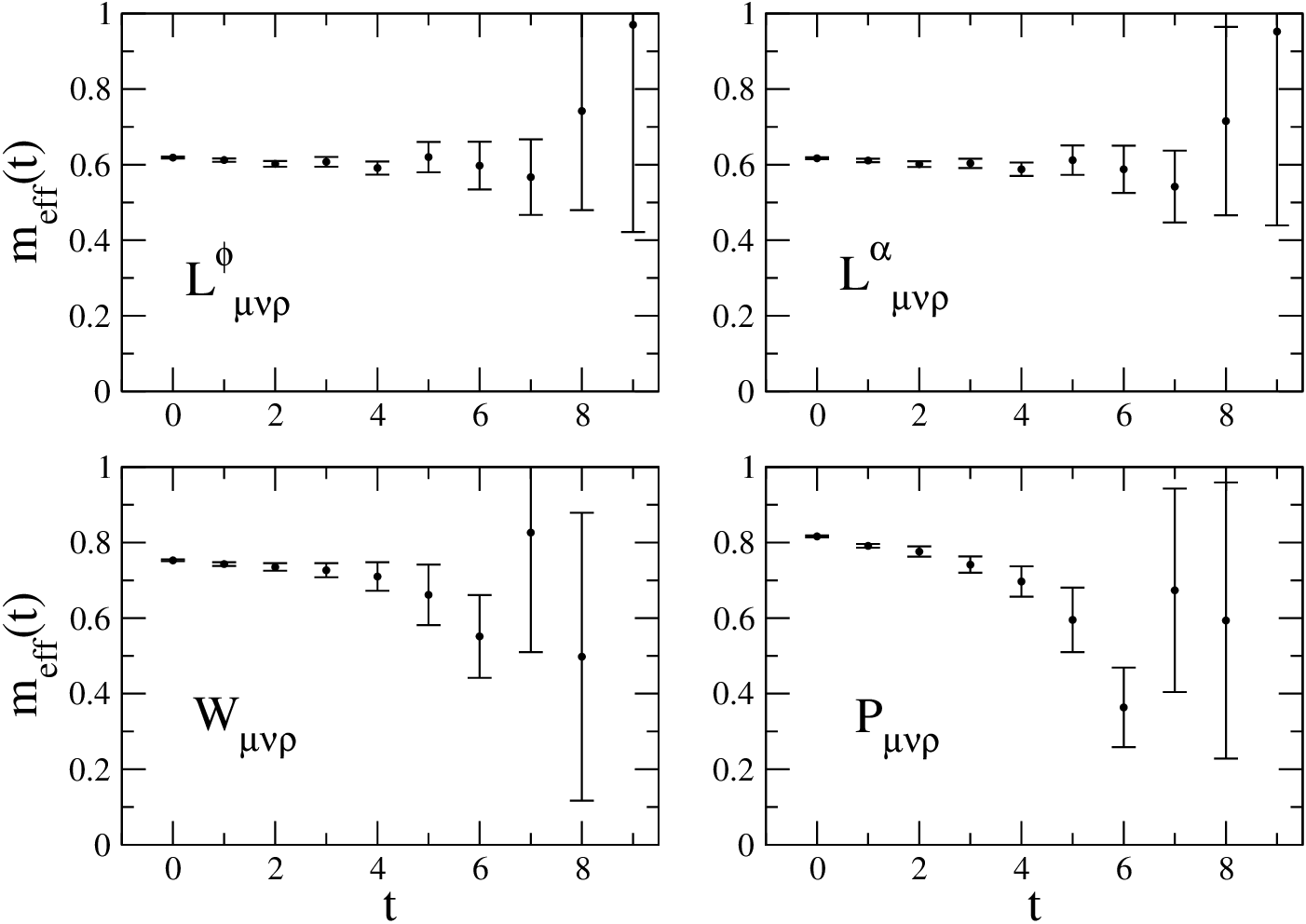}
\caption{Effective masses of the $I(\Lambda^P)=0(A_1^-)$ gauge-invariant link operators $L^\phi_{\mu\nu\rho}$ and $L^\alpha_{\mu\nu\rho}$, Wilson loop $W_{\mu\nu\rho}$ and Polyakov loop $P_{\mu\nu\rho}$ operators on a $20^3\times 40$ lattice with $\beta=8$, $\kappa=0.131$ and $\lambda=0.0033$.}
\label{graph_effmass_0A1-}
\end{figure}
To illustrate the efficacy of the operators, consider effective masses\footnote{In general one would use ${\cal O}^\dagger(0)$ rather than ${\cal O}(0)$, but in our SU(2) study the $\vec p=\vec 0$ operators are Hermitian and (as discussed in Sec.~\ref{Two-Particle Operators}) even the $\vec p\neq\vec 0$ correlation functions are statistically real.}
\begin{align}
m_\text{eff}(t) = - \log\left(\frac{\langle{\cal O}(t+1){\cal O}(0)\rangle}{\langle{\cal O}(t){\cal O}(0)\rangle}\right)
\end{align}
where ${\cal O}(t)$ is a gauge-invariant operator with its vacuum expectation value subtracted,
\begin{align}
{\cal O}(t) = O(t) - \left<O(t)\right>  \,\, . \label{subtractedO}
\end{align}
Figures~\ref{graph_effmass_0A1+} and~\ref{graph_effmass_0A1-} show effective mass plots for the $I(\Lambda^P)=0(A_1^+)$ and $0(A_1^-)$ channels of four operators: two gauge-invariant links, a Wilson loop, and a Polyakov loop.  The stout link and smearing parameters are $n_{\rm stout}=n_{\rm smear}=200$ and $r_{\rm stout}=r_{\rm smear}=0.1$.

For $0(A_1^+)$, the $L^\alpha_{\mu\nu\rho}$ and $P_{\mu\nu\rho}$ operators have nearly identical effective mass plots despite being conceptually very different operators.  The mass is near 0.4 in lattice units.
The $L^\phi_{\mu\nu\rho}$ operator with identical quantum numbers produces a different effective mass (near 0.3), and the $W_{\mu\nu\rho}$ operator gives another (noisier) result.
This is an indication that the $0(A_1^+)$ spectrum (corresponding to $J=0$ in the continuum) contains more than a lone Higgs boson.  A more sophisticated analysis method is presented in Sec.~\ref{sec:analysis} and applied in subsequent sections.

For $0(A_1^-)$, Fig.~\ref{graph_effmass_0A1-} provides four effective mass plots that collectively indicate a mass near 0.6 in lattice units.  Again this is $J=0$ in the continuum, and of course neither a single Higgs nor a single $W$ has $J^P=0^-$.  Our complete analysis of this and all other channels is discussed below.

\section{Correlation Matrix and Variational Method}\label{sec:analysis}

Particle energies, $E_n$, are extracted from lattice simulations by observing the exponential decay of correlation functions,
\begin{align}
C_{ij}(t) = \left< {\cal O}_i(t) {\cal O}_j(0) \right> &= \sum_n \left<0\right|{\cal O}_i\left|n\right> \left<n\right|{\cal O}_j\left|0\right> \exp\left(-E_n t \right) \\
&= \sum_n a_i^n a_j^n \exp\left(-E_n t \right)  \,\, ,
\end{align}
where ${\cal O}_i(t)$ is a Hermitian gauge-invariant operator with its vacuum expectation value subtracted as in Eq.~(\ref{subtractedO}).
The choice of operator determines the quantum numbers $I(\Lambda^P)$ of the states $\left|n\right>$ that are present in the correlation function and also determines the coupling strength, $a_i^n$, to each.  The operators are calculated for eight different levels of smearing, $n_{\rm stout}=n_{\rm smear}=0$, 5, 10, 25, 50, 100, 150, and 200.  The smearing parameters are held fixed at $r_{\rm stout}=r_{\rm smear}=0.1$.  Each of these different smearing levels produces a unique operator ${\cal O}_i$ in the correlation matrix $C_{ij}(t)$.

The energy spectrum is extracted using the variational method \cite{Kronfeld:1989tb,Luscher:1990ck}.  To begin, the eigenvectors $\vec{v}_n$ and eigenvalues $\lambda_n$ ($n=1,...,M$) of the correlation matrix are found at a single time step $C_{ij}(t_0)$ ($i,j=1,...,N$), where $N$ is the number of operators, $M$ is the number of statistically nonzero eigenvalues, which corresponds to the number of states that can be resolved, and $M\leq N$.  The value of $t_0$ is typically chosen to be small, e.g.\ $t_0=1$, where the signal-to-noise ratio is large.  The correlation matrix is changed from the operator basis to the eigenvector basis by
\begin{align}
\widetilde{C}_{nm}(t) = \frac{\vec{v}_n^T C(t) \vec{v}_m}{\sqrt{\lambda_n\lambda_m}} \,\, .  \label{eigenstate_correlation_matrix}
\end{align}
The correlation function for the $k\text{th}$ ($k=1,...,M$) state is then given by
\begin{align}
C_{k}(t) = \vec{R}_{k}^T \widetilde{C}(t) \vec{R}_{k} \,\, ,
\end{align}
where $\vec{R}_k$ is a set of orthonormal vectors chosen such that the energies from $C_k(t)$ are ordered from smallest to largest for increasing $k$.  $\vec{R}_{k}$ is determined recursively by a variational method as follows: $\vec{R}_1$ maximizes $C_1(t_1)$, the correlation function of the smallest energy at a time step $t_1>t_0$.  The normalization of Eq.~\eqref{eigenstate_correlation_matrix} ensures that $C_{k}(t_0)=1$, thus maximizing $C_1(t_1)$ ensures that $\vec{R}_1$ projects out the state with smallest energy while minimizing contamination from higher-energy states.  In practice, the time step $t_1$ is taken to be $t_0+1$.  The optimization of $C_1(t_1)$ reduces to solving the eigenproblem
\begin{align}
\widetilde{C}(t_1) \vec{x}_1 = \mu_1 \vec{x}_1 \,\, , \label{lagrange_multiplier}
\end{align}
where the eigenvalue $\mu_1$ is the Lagrange multiplier for the constraint $\vec{R}_1^T\vec{R}_1=1$, and the solution for $\vec{R}_1$ is given by the eigenvector $\vec{x}_1$ that maximizes $C_1(t_1)$.  The correlation function $C_2(t)$ of the next-smallest-energy state can be found by calculating $\vec{R}_2$ in the same way as above, given that $\vec{R}_2$ must be orthonormal to $\vec{R}_1$.  This is accomplished by defining $\vec{R}_2$ as the vector
\begin{align}
\vec{x}_2=\sum_{n=1}^{M-1} a_{n}\vec{x}_{1,n} \label{vec_x2}
\end{align}
that maximizes $C_2(t_1)$, where $\vec{R}_1=\vec{x}_{1,M}$ and $\vec{x}_{1,n}$ ($n=1,...,M-1$) are the remaining eigenvectors from Eq.~\eqref{lagrange_multiplier}.  The eigenproblem resulting from the maximization of $C_2(t_1)$ is
\begin{align}
X_1^T\widetilde{C}(t_1)X_1 \vec{a} = \mu_2 \vec{a} \,\, ,
\end{align}
where the matrix $X_1=(\vec{x}_{1,1},...,\vec{x}_{1,M-1})$, the vector $\vec{a}^T=(a_{1},...,a_{M-1})$ contains the coefficients from Eq.~\eqref{vec_x2} and the vector $\vec{R}_2~=~ X_1\vec{a}$ is calculated from the eigenvector $\vec{a}$ that maximizes $C_2(t_1)$.  The calculation can continue recursively up to the $M\text{th}$ case, where the eigenproblem becomes trivial.  The energy can then be extracted by a $\chi^2$-minimizing fit to a single exponential using
\begin{align}
C_{k}(t) = A_k \exp\left(-E_k t \right) \,\, .
\end{align}

\section{Spectrum at the physical point}\label{sec:spectrum}

\begin{figure}[tb]
\centering
\includegraphics[scale=0.6,clip=true]{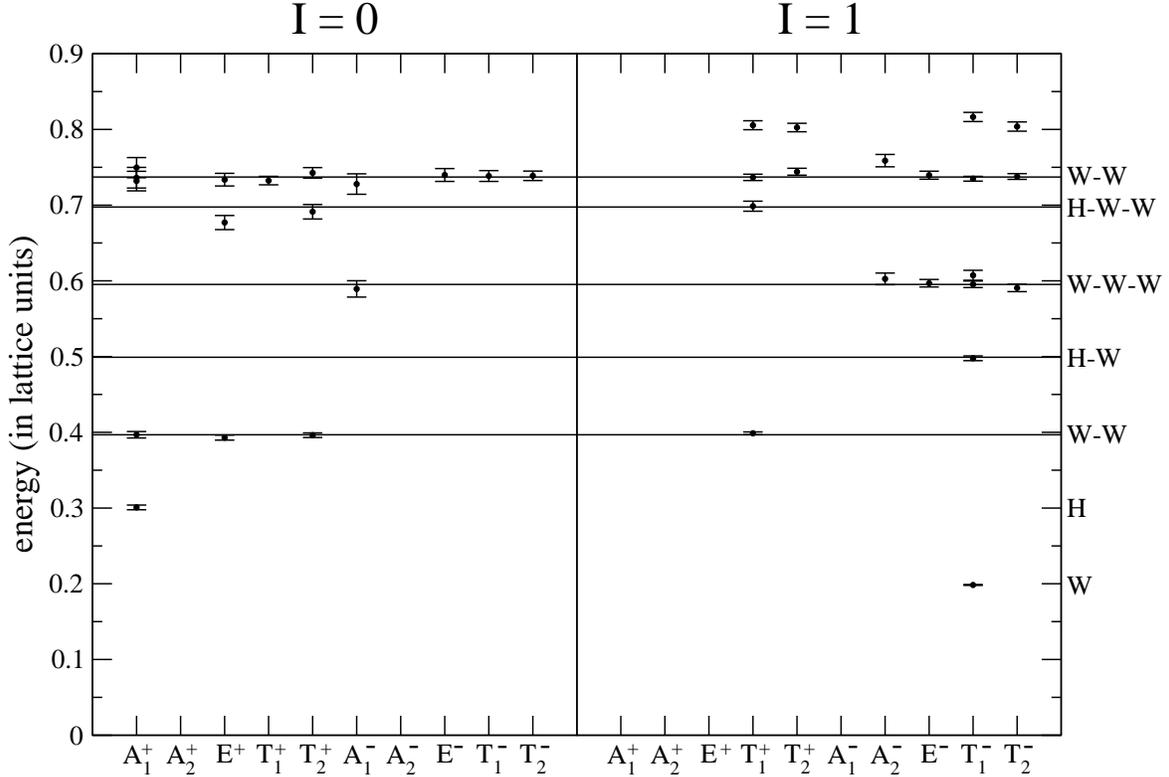}
\caption{Energy spectrum extracted from correlation functions of the gauge-invariant link operators $L^\phi_{\mu\nu\rho}$ and $L^\alpha_{\mu\nu\rho}$ for all isoscalar and isovector channels on a $20^3\times 40$ lattice with $\beta=8$, $\kappa=0.131$ and $\lambda=0.0033$.  These parameters put the theory very close to the experimental Higgs and $W$ boson masses.  Data points are lattice results with statistical errors; horizontal lines are the expectations from Eq.~\eqref{lattice_dispersion}.}
\label{graph_mass_link}
\end{figure}

\begin{figure}[tb]
\centering
\includegraphics[scale=0.6,clip=true]{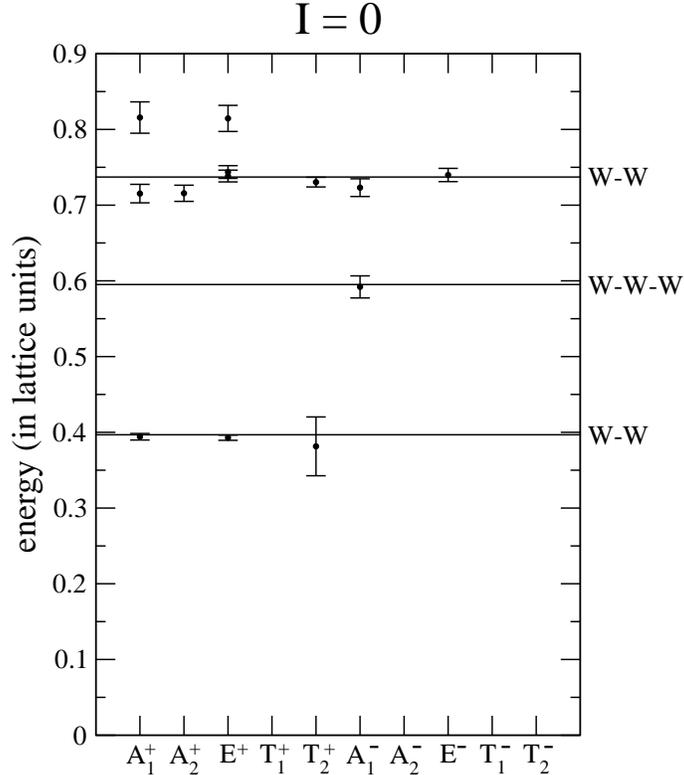}
\caption{Energy spectrum extracted from correlation functions of the Wilson loop and Polyakov loop operators $W_{\mu\nu\rho}$ and $P_{\mu\nu\rho}$ for all isoscalar channels on a $20^3\times 40$ lattice with $\beta=8$, $\kappa=0.131$ and $\lambda=0.0033$.  These parameters put the theory very close to the experimental Higgs and $W$ boson masses.  Data points are lattice results with statistical errors; horizontal lines are the expectations from Eq.~\eqref{lattice_dispersion}.}
\label{graph_mass_wilson_polyakov}
\end{figure}

Using the methods described above, an ensemble of 20,000 configurations was created on a
$20^3\times40$ lattice with $\beta=8$, $\lambda=0.0033$, and $\kappa=0.131$.
Figure \ref{graph_mass_link} shows the energy levels for isospins 0 and 1 as obtained from the gauge-invariant link operators $L^\phi_{\mu\nu\rho}$ and $L^\alpha_{\mu\nu\rho}$, and Fig.~\ref{graph_mass_wilson_polyakov} shows the energy levels for isospin 0 as obtained from the Wilson loop and Polyakov loop operators.
(Wilson/Polyakov loops cannot produce isospin 1, and lattice results for isospins higher than 1 are not considered in this work.)
As expected, the lightest state in the spectrum has $I(\Lambda^P)=1(T_1^-)$ corresponding to
a single $W$ boson.  The mass is near 0.2 in lattice units (with a tiny statistical error)
and identification with the experimentally known $W$ mass allows us to infer the lattice
spacing in physical units.

The next energy level above the single $W$ has an energy near 0.3 and is observed in the
$0(A_1^+)$ channel, exactly as expected for the Higgs boson.
Our lattice parameters were tuned to put this mass near its
experimental value; the result from our simulation is $122\pm1$ GeV.
Notice that neither the single $W$ nor the single Higgs is observed from the Wilson loop or
Polyakov loop, but both are seen from the gauge-invariant link operators.
Moreover, notice that the Higgs boson $H$ has not been created by just a single
$\phi(x)$ but rather by gauge-invariant operators that
can never contain any odd power of $\phi(x)$.
Much like QCD, physical particles in the observed spectrum do not present any
obvious linear one-to-one correspondence with fields in the Lagrangian.
For a recent discussion in the context of a gauge-fixed lattice study, see
Refs.~\cite{Maas:2012tj,Maas:2012zf}.

Continuing upward in energy within Figs.~\ref{graph_mass_link} and \ref{graph_mass_wilson_polyakov},
we see a signal with energy at $2m_W$ in four specific channels.
These are exactly the four channels that correspond to the allowed quantum numbers of
a pair of {\em stationary} $W$ bosons.  In the continuum, the wave function for such a pair
of spin-1 $W$ bosons would be the product of a spin part and an isospin part.  The total
wave function must be symmetric under particle interchange.  This permits just two
continuum states with isospin 0 [$0(0^+)$ and $0(2^+)$], and a single continuum
state with isospin 1 [$1(1^+)$].
Note that the parity of a $W$ pair is always positive in the absence of orbital angular momentum.
A glance at Table~\ref{irrep_spin_table} reveals that these continuum states
match the lattice observations at energy $2m_W$ perfectly.
An energy shift away from $2m_W$ would represent binding energy or a scattering state,
but no shift is visible in our lattice simulation at this weak coupling value.

The next state in Figs.~\ref{graph_mass_link} and \ref{graph_mass_wilson_polyakov}
has an energy of $m_H+m_W$ and is another pair of stationary bosons.  Because the Higgs boson
is $0(0^+)$, the Higgs-$W$ pair should have the quantum numbers of the $W$.  The lattice
data show that the Higgs-$W$ pair does indeed appear in exactly the same $I(\Lambda^P)$ channels
as does the single $W$.

Two states are expected to appear with an energy near 0.6 because this corresponds to
$2m_H\approx3m_W$.  A pair of stationary Higgs bosons should have the same quantum numbers
as a single Higgs, i.e.\ $I(J^P)=0(0^+)$, but no such signal appears in
Figs.~\ref{graph_mass_link} and \ref{graph_mass_wilson_polyakov}.
To see this two-Higgs state we will need a different creation operator;
Sec.~\ref{Two-Particle Operators} introduces this operator and uses it
to observe the two-Higgs state within our lattice simulations.

A collection of three stationary $W$ bosons must have a wave function that is symmetric under
interchange of any pair, and must be built from a spin part and an isospin part.
The $I=0$ case has an antisymmetric isospin part and the only available
antisymmetric spin part is $J=0$.  The $I=1$ case is of mixed symmetry and can combine
with $J=1$, 2, or 3 (but not $J=0$) to form a symmetric wave function.
These continuum options, i.e.\ $0(0^-)$, $1(1^-)$, $1(2^-)$ and $1(3^-)$, can be
converted into lattice channels easily by using Table~\ref{irrep_spin_table} and
the result is precisely the list of channels observed in
Figs.~\ref{graph_mass_link} and \ref{graph_mass_wilson_polyakov}, i.e.\
$0(A_1^-)$, $1(T_1^-)$, $1(E^-)$, $1(T_2^-)$, and $1(A_2^-)$.

The next energy level is $m_H+2m_W$ which should have identical $I(\Lambda^P)$ options
to the pair of stationary $W$ bosons discussed above.  Figure~\ref{graph_mass_link}
verifies this expectation, having signals for $0(A_1^+)$, $0(E^+)$, $0(T_2^+)$, and
$1(T_1^+)$, although errors bars are somewhat larger for this high energy state.

The next energy level in
Figs.~\ref{graph_mass_link} and \ref{graph_mass_wilson_polyakov}
is a pair of {\em moving} $W$ bosons with vanishing {\em total} momentum.
Recall that our operators were defined to have zero total momentum, but this still
permits a two-particle state
where the particles have equal and opposite momenta.
Momentum components along the $x$, $y$ or $z$ axes of the lattice can have integer multiple values of $2\pi/L$, where $L$ is the spatial length of the lattice.  The lattice dispersion relation for a boson with mass $m$ and momentum $\vec{p}$ is 
\begin{align}
\sinh^2\left(\frac{aE(\vec{p})}{2}\right) = \sinh^2\left(\frac{am}{2}\right) + \sum_{i=1}^3 \sin^2\left(\frac{ap_i}{2}\right) \label{lattice_dispersion}
\end{align}
which reduces to the continuum relation,
$E(\vec{p}) = \sqrt{m^2 + \vec{p}^2}$,
as the lattice spacing $a$ goes to zero.  Given the lattice spacing and statistical precision used in this paper, the difference between Eq.~\eqref{lattice_dispersion} and the continuum relation is noticeable.  The energy of a state of two noninteracting bosons is simply $E_1(\vec{p}_1)+E_2(\vec{p}_2)$, with energies from Eq.~\eqref{lattice_dispersion}.

Two particles with relative motion can also have orbital angular momentum $L$;
the allowed $I(J^P)$ for Higgs-Higgs, Higgs-$W$ and $W$-$W$ states are listed in Table~\ref{orbital_angular_momentum}.
\begin{table}[tb]
\caption{$I(J^P)$ quantum numbers for Higgs-Higgs, Higgs-$W$ and $W$-$W$ states with orbital angular momentum $L$. Higgs-Higgs states must have positive parity due to Bose statistics.}
\label{orbital_angular_momentum}
\centering
\begin{tabular}{c|c|c|c|c} 
 &    Higgs-Higgs    &            Higgs-$W$             &    \multicolumn{2}{c}{$W$-$W$} \\
\hline
~$L$~ &    $I=0$     &            $I=1$              &    $I=0$     &            $I=1$ \\
\hline
0 & $0^+$ &          $1^-$          & $0^+$, $2^+$ &          $1^{+}$ \\
1 &  ---  & $0^{+}$, $1^+$, $2^{+}$ & $1^-$, $2^-$, $3^-$ & $0^{-}$, $1^{-}$, $2^{-}$ \\
2 & $2^+$ & $1^{-}$, $2^-$, $3^{-}$ & $0^+$, $1^+$, $2^+$, $3^+$, $4^+$ & $1^{+}$, $2^{+}$, $3^{+}$ \\
3 &  ---  & $2^{+}$, $3^+$, $4^{+}$ & $1^-$, $2^-$, $3^-$, $4^-$, $5^-$ & $2^{-}$, $3^{-}$, $4^{-}$ \\
\vdots & \vdots & \vdots & \vdots & \vdots \\
\end{tabular} 
\end{table}
There is no way to specify $L$ with lattice operators because it is not a conserved quantum number; only the total momentum $J$ can be specified, which corresponds to $\Lambda$ in a lattice simulation.
For two moving $W$ particles, all quantum numbers with $I=0$ or 1 are possible except $0(0^-)$ and $1(0^+)$.  Therefore a signal could appear in all $I(\Lambda^P)$ channels, even $0(A_1^-)$ and $1(A_1^+)$ because of $J=4$ states.
As evident from Figs.~\ref{graph_mass_link} and \ref{graph_mass_wilson_polyakov},
our lattice simulation produced signals in many channels, but not in all.
Section~\ref{Two-Particle Operators} provides the
explanation for why this particular subset of channels did not show a signal.

Beyond this large energy, we are approaching the limit of the reach of this set of operators.
A few data points are shown at even higher energies (in the neighborhood of $4m_W$) in Figs.~\ref{graph_mass_link} and \ref{graph_mass_wilson_polyakov},
but a confident interpretation of those will require further computational effort that is presented in Secs.~\ref{sec:biggerlattice} and \ref{sec:infiniteHiggs}.

To conclude this section, it is interesting to notice a clear qualitative distinction
between the Wilson/Polyakov loop operators and the gauge-invariant link operators:
the former (Fig.~\ref{graph_mass_wilson_polyakov}) found only pure $W$ boson states
whereas the latter (Fig.~\ref{graph_mass_link}) found additional states containing one
Higgs boson.  States containing two Higgs bosons must wait until Sec.~\ref{Two-Particle Operators}.

\section{Spectrum on a larger lattice}\label{sec:biggerlattice}

To confirm that several of the states in Figs.~\ref{graph_mass_link} and \ref{graph_mass_wilson_polyakov} are truly multiparticle states with linear momentum, the simulations of the previous section are repeated using a larger lattice volume.  Since momentum on a lattice is given by integer multiples of $2\pi/L$, where $L$ is the spatial length of the lattice, increasing the lattice volume should cause the energies of states with linear momentum to decrease by a predictable amount.  Here the lattice parameters are set to $\beta = 8$, $\lambda = 0.0033$, $\kappa = 0.131$, which is the same as the previous section, but now the lattice volume is $24^3\times 48$.  An ensemble of 20,000 configurations is used.

\begin{figure}[tb]
\centering
\includegraphics[scale=0.6,clip=true]{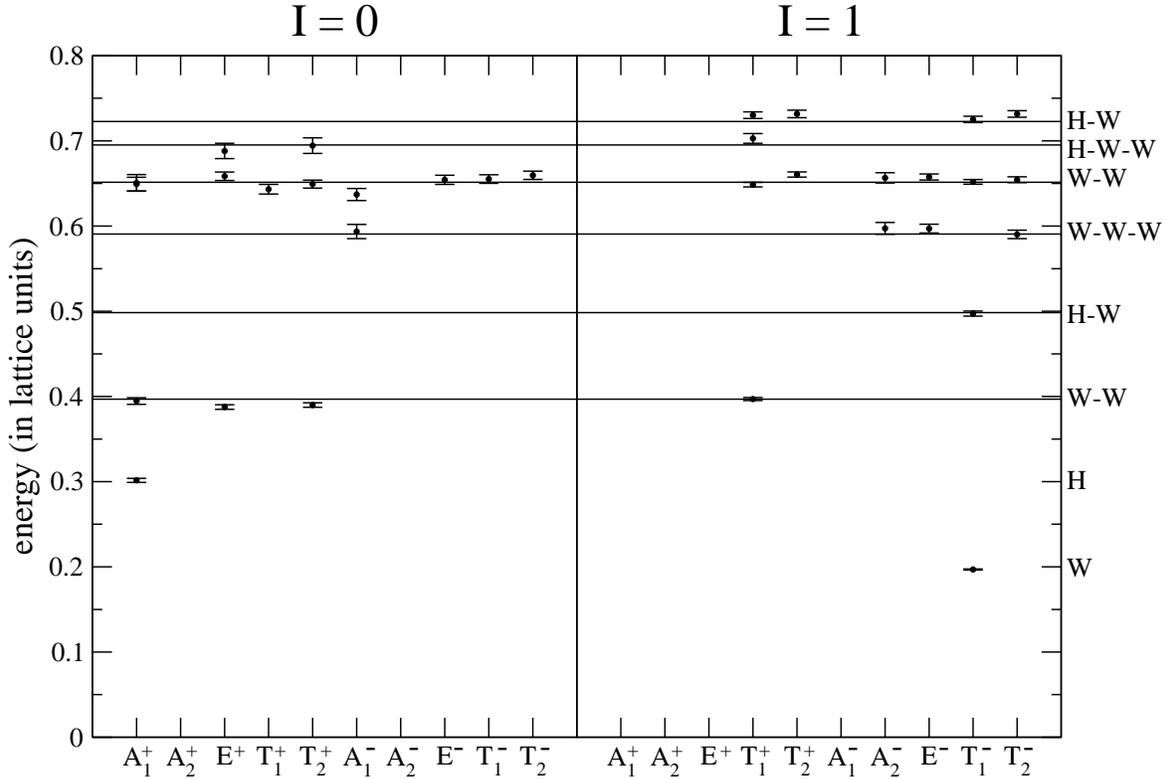}
\caption{The same as Fig.~\ref{graph_mass_link} but using a $24^3\times 48$ lattice.}
\label{graph_mass_link_24x24x24x48}
\end{figure}
\begin{figure}[tb]
\centering
\includegraphics[scale=0.6,clip=true]{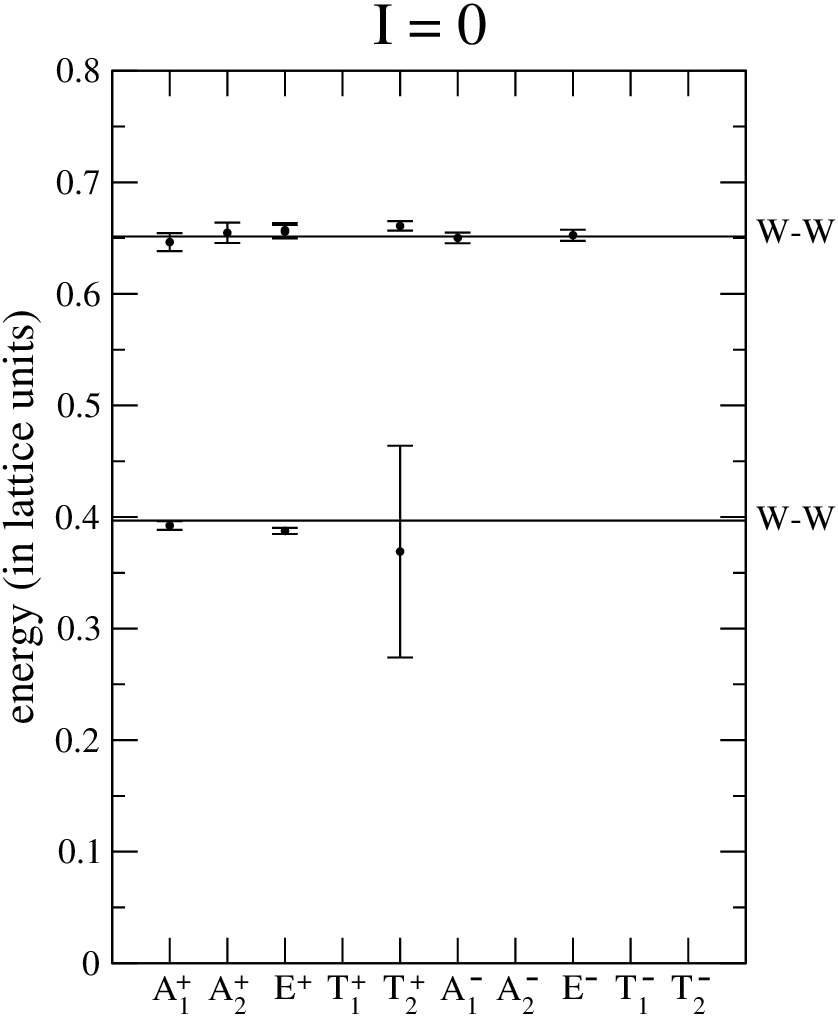}
\caption{The same as Fig.~\ref{graph_mass_wilson_polyakov} but using a $24^3\times 48$ lattice.}
\label{graph_mass_wilson_polyakov_24x24x24x48}
\end{figure}

The energy spectrum, extracted by a variational analysis, is shown in Figs.~\ref{graph_mass_link_24x24x24x48} and \ref{graph_mass_wilson_polyakov_24x24x24x48}.  The Higgs and $W$ masses remain virtually unchanged, with a Higgs mass of $123\pm 1$ GeV.  This stability indicates that finite volume artifacts are negligible.

The data points that lie at 0.65 in lattice units correspond perfectly to two $W$ particles with the minimal nonzero linear momentum.  This physics appears in Figs.~\ref{graph_mass_link} and \ref{graph_mass_wilson_polyakov} at a larger energy, and the energy shift is in numerical agreement with the change in energy due to changing the lattice volume.
Also, the four data points at 0.8 in Fig.~\ref{graph_mass_link} were numerically compatible with (a) a Higgs-$W$ pair moving back-to-back with the minimal momentum or (b) a collection of four $W$ bosons all at rest.  This physics has energy 0.73 in Fig.~\ref{graph_mass_link_24x24x24x48} which cannot be a four-$W$ state but is in good agreement with a back-to-back Higgs-$W$ pair.
From Table~\ref{orbital_angular_momentum} all $J^P$ quantum numbers except $0^-$ are allowed for a moving Higgs-$W$ pair, but these lattice operators have found a signal in only a few channels.
Section~\ref{Two-Particle Operators} addresses the issue of missing irreducible representations for multiparticle states with momentum.

It is noteworthy that some states consisting of three stationary $W$ particles, $1(T_1^-)$ in Fig.~\ref{graph_mass_link_24x24x24x48} and $0(A_1^-)$ in Fig.~\ref{graph_mass_wilson_polyakov_24x24x24x48}, as well as the $0(A_1^+)$ Higgs-$W$-$W$ state in Fig.~\ref{graph_mass_link_24x24x24x48}, were not detected in the larger lattice volume.  This is because the variational analysis cannot resolve these states from the current basis of operators.  When the lattice volume was increased, the spectral density increased as more multiparticle states became detectable in the correlation functions.  As a result, states with a small overlap with the basis of operators could not be successfully extracted, even though they had been observed for the smaller lattice volume.  Of course, these states could be seen again if the basis of operators was improved, for example, by increasing the number of operators.

\section{Spectrum with a heavy Higgs}
\label{sec:infiniteHiggs}

A simple method to confirm which of the multiparticle states in Figs.~\ref{graph_mass_link} and \ref{graph_mass_wilson_polyakov} contain a Higgs boson is to change the Higgs mass and leave everything else unchanged.  Here we choose the extreme case of an infinite quartic coupling, corresponding to the maximal Higgs mass \cite{Hasenfratz:1987uc,Langguth:1987vf,Hasenfratz:1987eh}.  The lattice parameters are set to $\beta = 8$, $\lambda = \infty$, $\kappa=0.40$, and the geometry is $20^3\times40$.  An ensemble of 20,000 configurations is used.  With these parameters, the $W$ mass in lattice units is nearly identical to the value in Fig.~\ref{graph_mass_link}.

\begin{figure}[tb]
\centering
\includegraphics[scale=0.6,clip=true]{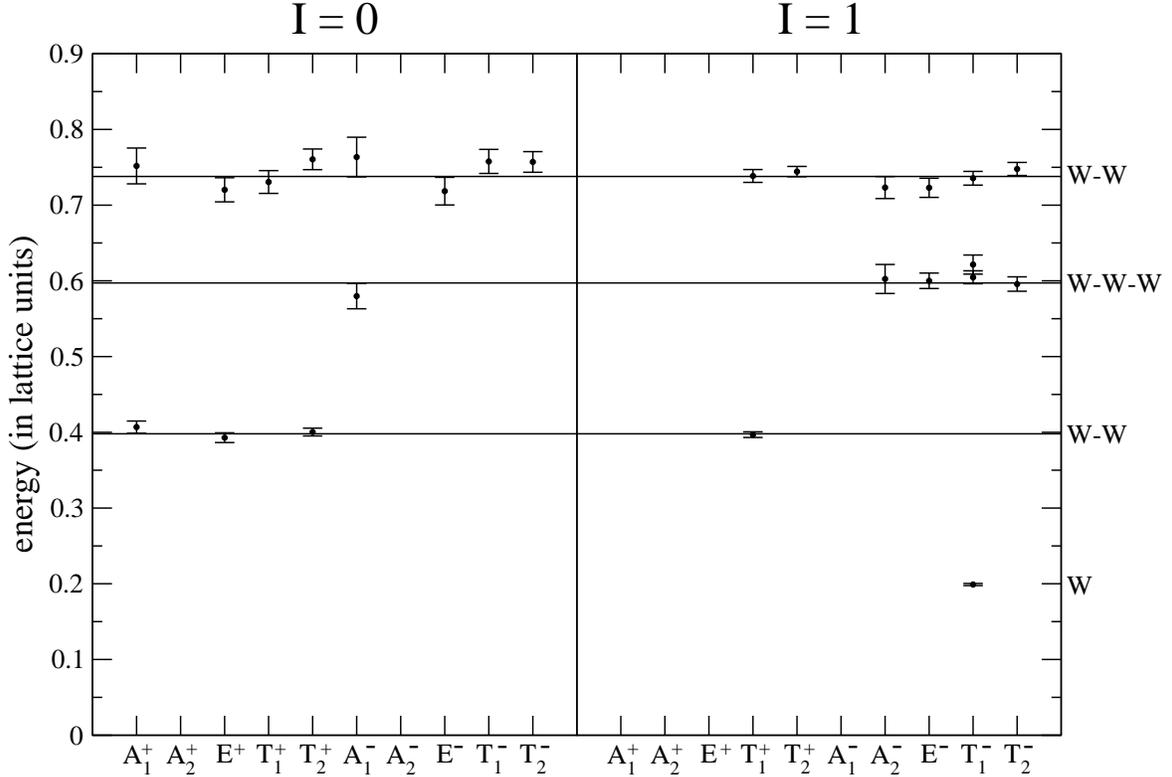}
\caption{The same as Fig.~\ref{graph_mass_link} but using $\kappa=0.40$ and $\lambda=\infty$.  The Higgs mass is off the graph because of its large value.}
\label{graph_mass_link_lambda_inf}
\end{figure}

\begin{figure}[tb]
\centering
\includegraphics[scale=0.6,clip=true]{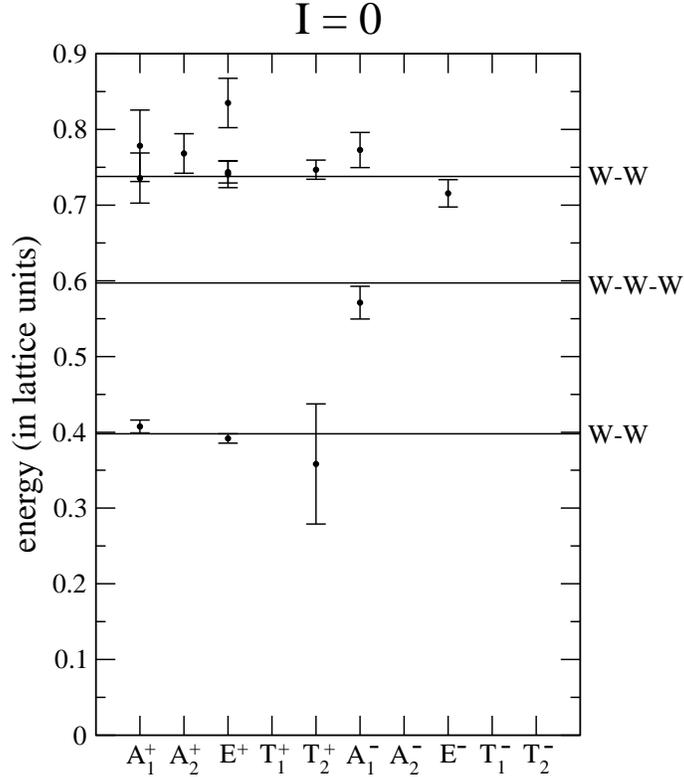}
\caption{The same as Fig.~\ref{graph_mass_wilson_polyakov} but using $\kappa=0.40$ and $\lambda=\infty$.}
\label{graph_mass_wilson_polyakov_lambda_inf}
\end{figure}

The energy spectrum, extracted by a variational analysis as usual, is shown in Figs.~\ref{graph_mass_link_lambda_inf} and \ref{graph_mass_wilson_polyakov_lambda_inf}.
The spectrum of states containing $W$ particles remains essentially the same as in Figs.~\ref{graph_mass_link} and \ref{graph_mass_wilson_polyakov} but all states with Higgs content are no longer visible.
This is consistent with the notion that the Higgs mass is now so large that all states with Higgs content have been pushed up to a higher energy scale.

To test this expectation of a large Higgs mass, a simultaneous fit of the entire $0(A_1^+)$ gauge-invariant-link correlation matrix was performed.
(For a comparison of this method to the variational analysis in a different
lattice context, see Ref.~\cite{Lewis:2011ti}.)
A three-state fit to time steps $t\ge2$ provided a good description of the lattice data,
with a $\chi^2/\text{d.o.f.}=0.84$.
The smallest energy corresponds to a pair of stationary $W$ bosons,
the next energy is a pair of $W$ bosons moving back-to-back with vanishing total momentum,
and the third energy is $1.8\pm0.2$ in lattice units which is $720\pm70$~GeV.
This third energy is consistent with the maximal Higgs energy found in early lattice studies \cite{Hasenfratz:1987uc,Langguth:1987vf,Hasenfratz:1987eh}.
Lattice artifacts will be significant
for this Higgs mass, since it is larger than unity in lattice units.
For our purposes it is sufficient to conclude that the Higgs mass is much larger than
the low-lying spectrum of multiparticle $W$-boson states.
This study of the spectrum in a heavy-Higgs world reinforces our understanding of which
states in the spectrum contain a Higgs boson.

\section{Two-Particle Operators}
\label{Two-Particle Operators}

The operators used in previous sections of this work were, at most, quadratic in the field $\phi(x)$.
They led to excellent results for several states in the SU(2)-Higgs spectrum, including
multiboson states, but additional operators can accomplish even more.
In particular, recall that the two-Higgs state was not observed in previous sections, the two-$W$ state with internal linear momentum was missing from a few $I(\Lambda^P)$ channels, and the Higgs-$W$ state with internal linear momentum was similarly missing from some $I(\Lambda^P)$ channels.

Presently, multiparticle operators will be constructed and the allowed irreducible representations will be compared to the results in Figs.~\ref{graph_mass_link} and \ref{graph_mass_wilson_polyakov}.  A two-particle operator ${\cal O}^{AB}(t)$ can be obtained by multiplying two operators with the following vacuum subtractions:
\begin{align}
{\cal O}^{AB}(t) &= {\cal O}^{A}(t) {\cal O}^{B}(t) - \left<{\cal O}^{A}(t) {\cal O}^{B}(t)\right> \,\, , \\
{\cal O}^{A}(t) &= {O}^{A}(t) - \left<{O}^{A}(t)\right> \,\, , \\
{\cal O}^{B}(t) &= {O}^{B}(t) - \left<{O}^{B}(t)\right> \,\, ,
\end{align}
where ${\cal O}^{A}(t)$ and ${\cal O}^{B}(t)$ each couple predominantly to a single-particle state.  The two-particle correlation function is then simply
\begin{align}
C^{AB}(t) = \left<{\cal O}^{AB}(t) {\cal O}^{AB\dag}(0)\right> \,\, .
\end{align}
Note that ${\cal O}^{AB}(t)$ is not strictly a two-particle operator because all states with the same quantum numbers as ${\cal O}^{AB}(t)$ can be created by it, including single-particle states.  However, this construction will result in a much stronger overlap with the two-particle states, such as Higgs-Higgs which was not found using the operators in Sec.~\ref{sec:operators}.  A three-particle operator is defined similarly:
\begin{align}
{\cal O}^{ABC}(t) &= {\cal O}^{A}(t) {\cal O}^{B}(t) {\cal O}^{C}(t) - \left<{\cal O}^{A}(t) {\cal O}^{B}(t) {\cal O}^{C}(t)\right> \,\, .
\end{align}

In this section we have written the correlation function using the Hermitian conjugate because we intend to use operators with nonzero momentum, whereas in the previous sections all operators were strictly Hermitian.  This does not affect our variational method because all of our correlation functions are real; to be precise, the imaginary component of each correlation function is equal to zero within statistical fluctuations.

The single-particle operators for the Higgs and $W$ are given by
\begin{align}
H(\vec{p}) &=  \sum_{\vec{x}} \frac{1}{2} \operatorname{Tr}\left\{\phi^\dag(x)\phi(x)\right\} \, \exp\left\{i\vec{p}\cdot\vec{x}\right\} \,\, , \label{Hp} \\
W^a_{\mu}(\vec{p}) &= \sum_{\vec{x}} \frac{1}{2} \operatorname{Tr}\left\{-i\sigma^a \phi^\dag(x) U_\mu(x) \phi(x+\hat{\mu})\right\} \, \exp\left\{i\vec{p}\cdot\left(\vec{x}+\tfrac{1}{2}\hat{\mu}\right)\right\} \,\, , \label{Wp}
\end {align}
where $\vec{p}$ is the momentum and has components given by integer multiples of $2\pi/L$ in the $x$, $y$ or $z$ directions, with $L$ being the spatial length of the lattice.  Combining the $W$ operators requires some additional care due to the isospin indices.  $W$-$W$ eigenstates of $I$ are obtained using the scalar and vector products
\begin{align}
&I=0: \quad \vec{W}_\mu\cdot\vec{W}_\nu = W_\mu^aW_\nu^a \,\, , \\
&I=1: \quad \vec{W}_\mu\times\vec{W}_\nu = \epsilon^{abc}W_\mu^bW_\nu^c \,\, ,
\end{align}
where the repeated $a$, $b$, $c$ indices are summed.  Combinations of $W$ operators with $I>1$ are not considered in this paper.  The irreducible representations of the $W$-$W$ operators with $\vec{p}=\vec0$ are given by
\begin{align}
0(A_1^{+}) :& \quad W_1^aW_1^a+W_2^aW_2^a+W_3^aW_3^a \\
0(E^{+})   :&\quad \frac{W_1^aW_1^a-W_2^aW_2^a}{\sqrt{2}}, \frac{W_1^aW_1^a+W_2^aW_2^a-2W_3^aW_3^a}{\sqrt{6}}\\
0(T_2^{+}) :&\quad W_1^aW_2^a,W_2^aW_3^a,W_3^aW_1^a \\
1(T_1^{+})  :&\quad \epsilon^{abc}W_1^bW_2^c,\epsilon^{abc}W_2^bW_3^c,\epsilon^{abc}W_3^bW_1^c
\end{align}
which correspond to the allowed continuum spins.  The isospin combinations for three $W$'s with $I=0$ or $1$ are
\begin{align}
&I=0; \quad \vec{W}_\mu\cdot\left(\vec{W}_\nu\times\vec{W}_\rho\right) = \epsilon^{abc}W_\mu^aW_\nu^bW_\rho^c  \,\, , \\
&I=1: \quad \vec{W}_\mu\left(\vec{W}_\nu\cdot\vec{W}_\rho\right) = W_\mu^aW_\nu^bW_\rho^b \,\, , \\
&I=1: \quad \vec{W}_\mu\times\left(\vec{W}_\nu\times\vec{W}_\rho\right) = \epsilon^{abc}\epsilon^{cde}W_\mu^bW_\nu^dW_\rho^e \,\, .
\end{align}
(Unnecessary for our purposes is another $I=1$ triple-$W$ operator, formed by combining an $I=2$ pair with the third $W$.)

\begin{table}[tb]
\caption{Octahedral group multiplicities of Higgs-Higgs, Higgs-$W$, $W$-$W$ and $W$-$W$-$W$ operators built of the operators in Eqs.~(\ref{Hp}) and (\ref{Wp}) with $\vec p=\vec 0$.  Repeated SU(2) indices $a$, $b$, $c$ are summed, but Lorentz indices $\mu$, $\nu$, $\rho$ are not.  The indices $\mu$, $\nu$, $\rho$ are not equal to one another.}
\label{zero_momentum_multiplicities}
\begin{tabular}{r|c|ccccccccccc}
Operator                                              &$~I~$&$A_1^+$&$A_2^+$&$E^+$&$T_1^+$&$T_2^+$&$A_1^-$&$A_2^-$&$E^-$&$T_1^-$&$T_2^-$\\
\hline 
$HH$                                                  &  0  &   1   &   0   &  0  &   0   &   0   &   0   &   0   &  0  &   0   &   0   \\
\hline 
$HW_\mu^a$                                            &  1  &   0   &   0   &  0  &   0   &   0   &   0   &   0   &  0  &   1   &   0   \\
\hline
$W_\mu^aW_\mu^a$                                      &  0  &   1   &   0   &  1  &   0   &   0   &   0   &   0   &  0  &   0   &   0   \\
$W_\mu^aW_\nu^a$                                      &  0  &   0   &   0   &  0  &   0   &   1   &   0   &   0   &  0  &   0   &   0   \\
$\epsilon^{abc}W_\mu^bW_\nu^c$                        &  1  &   0   &   0   &  0  &   1   &   0   &   0   &   0   &  0  &   0   &   0   \\
\hline 
$\epsilon^{abc}W_\mu^aW_\nu^bW_\rho^c$                &  0  &   0   &   0   &  0  &   0   &   0   &   1   &   0   &  0  &   0   &   0   \\
$W_\mu^aW_\mu^bW_\mu^b$                               &  1  &   0   &   0   &  0  &   0   &   0   &   0   &   0   &  0  &   1   &   0   \\
$W_\mu^aW_\mu^bW_\nu^b$                               &  1  &   0   &   0   &  0  &   0   &   0   &   0   &   0   &  0  &   1   &   1   \\
$W_\mu^aW_\mu^bW_\mu^b$                               &  1  &   0   &   0   &  0  &   0   &   0   &   0   &   0   &  0  &   1   &   1   \\
$W_\mu^aW_\nu^bW_\rho^b$                              &  1  &   0   &   0   &  0  &   0   &   0   &   0   &   1   &  1  &   0   &   0   \\
$\epsilon^{abc}\epsilon^{cde}W_\mu^bW_\mu^dW_\nu^e$   &  1  &   0   &   0   &  0  &   0   &   0   &   0   &   0   &  0  &   1   &   1   \\
$\epsilon^{abc}\epsilon^{cde}W_\mu^bW_\nu^dW_\rho^e$  &  1  &   0   &   0   &  0  &   0   &   0   &   0   &   0   &  1  &   0   &   0   \\
\end{tabular} 
\end{table}

\begin{figure}[tb]
\centering
\includegraphics[scale=0.6,clip=true]{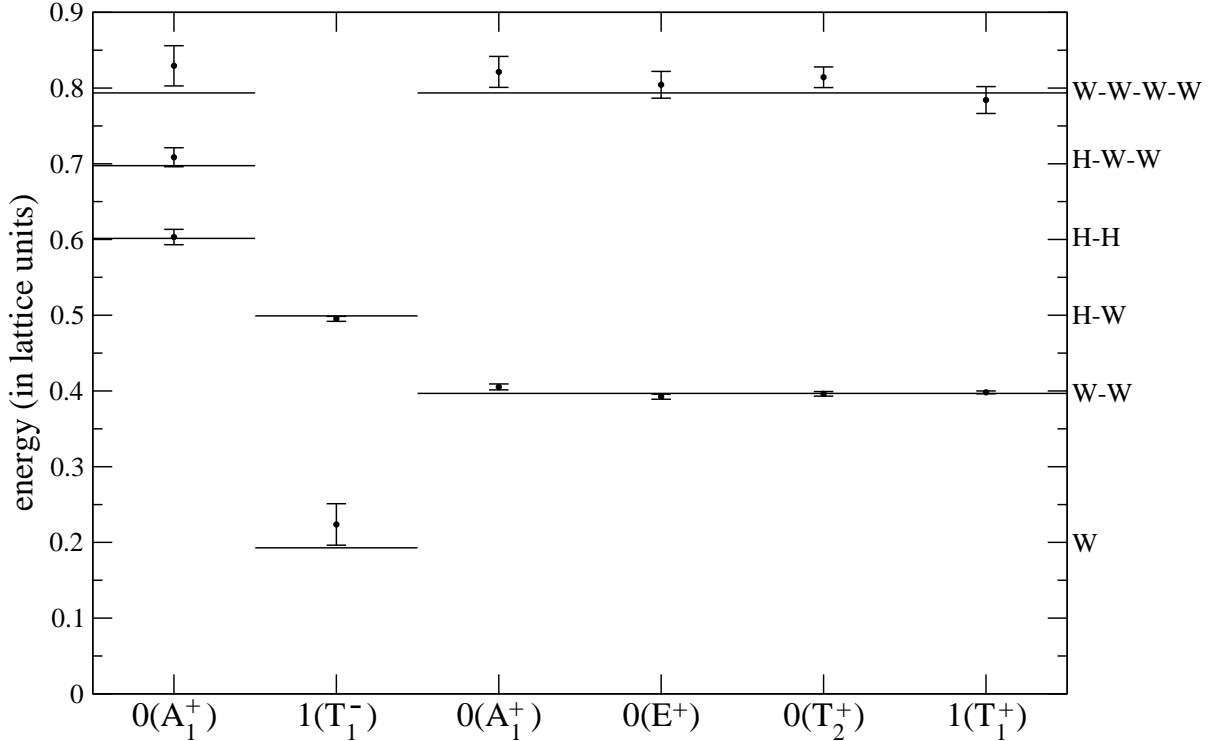}
\caption{Energy spectrum extracted from correlation functions of Higgs-Higgs, Higgs-$W$ and $W$-$W$ operators built from Eqs.~(\ref{Hp}) and (\ref{Wp}) with $\vec p=\vec 0$ on a $20^3\times 40$ lattice with $\beta=8$, $\kappa=0.131$ and $\lambda=0.0033$.  Data points are lattice results with statistical errors; horizontal lines are the expectations from Eq.~\eqref{lattice_dispersion}.}
\label{graph_mass_hh_hw_ww}
\end{figure}

Table~\ref{zero_momentum_multiplicities} shows the multiplicities for Higgs-Higgs, Higgs-$W$, $W$-$W$ and $W$-$W$-$W$ operators built entirely of $\vec p=\vec 0$ operators.
The energy spectrum obtained from the two-boson operators by variational analysis is displayed
in Fig.~\ref{graph_mass_hh_hw_ww}.
The two-Higgs state, absent until now, is seen quite precisely.  The $W$-$W$ and Higgs-$W$ signals
are also excellent.  Even three-boson and four-boson states are observed.
(Readers of Sec.~\ref{sec:biggerlattice} might wonder whether the four-$W$ states in
Fig.~\ref{graph_mass_hh_hw_ww} could instead be a Higgs-$W$ state with momentum.
Recall, though, that a Higgs-$W$ state cannot have isospin 0.)
Another success worth noticing is that the single Higgs does not appear at all
and the single $W$ couples only weakly; that is a success because the operators were
intended to be multiparticle operators.

\begin{figure}[tb]
\centering
\includegraphics[scale=0.6,clip=true]{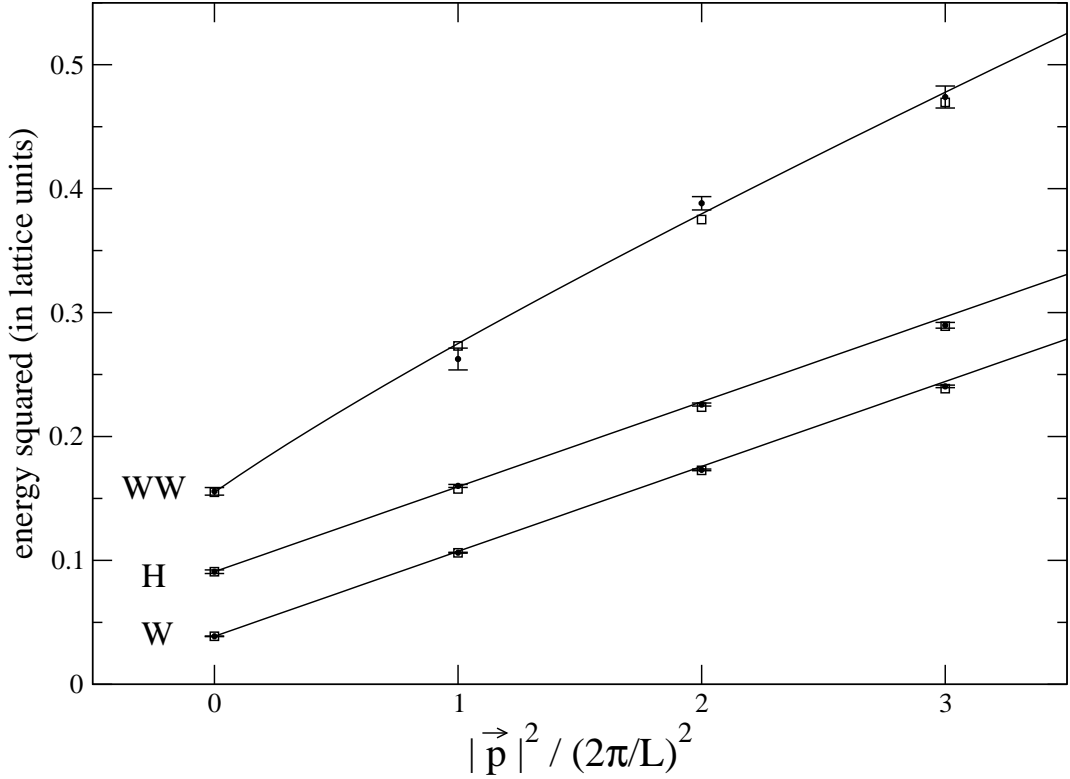}
\caption{Energy spectrum extracted from correlation functions of $H(\vec{p})$ and $W^a(\vec{p})$ operators from Eqs.~(\ref{Hp}) and (\ref{Wp}) as a function of momentum $\vec{p}$ on a $24^3\times 48$ lattice with $\beta=8$, $\kappa=0.131$ and $\lambda=0.0033$.  Data points are lattice results with statistical errors; solid curves are based on the continuum dispersion relation $E^2=m^2+\vec{p}^2$; empty boxes are the expectations from the lattice dispersion relation Eq.~\eqref{lattice_dispersion}.}
\label{graph_mass_h_w_momentum}
\end{figure}

The operators $H(\vec{p})$ and $W^a_\mu(\vec{p})$ from Eqs.~\eqref{Hp} and \eqref{Wp} were calculated for momenta given by $\left|\vec{p}\right|=2\pi/L$, $\left|\vec{p}\right|=\sqrt{2}(2\pi/L)$ and $\left|\vec{p}\right|=\sqrt{3}(2\pi/L)$.
Figure~\ref{graph_mass_h_w_momentum} shows the spectrum obtained from a variational analysis of the single Higgs and $W$ operators versus momentum.
Both Higgs and $W$ operators contain an excited state which is a two-$W$ state, where one $W$ is stationary and the other has momentum.  Notice that the two-$W$ energy does not form a straight line since its continuum relation is $E = m + \sqrt{m^2+\vec{p}^2}$.

\begin{figure}[tb]
\centering
\includegraphics[scale=0.6,clip=true]{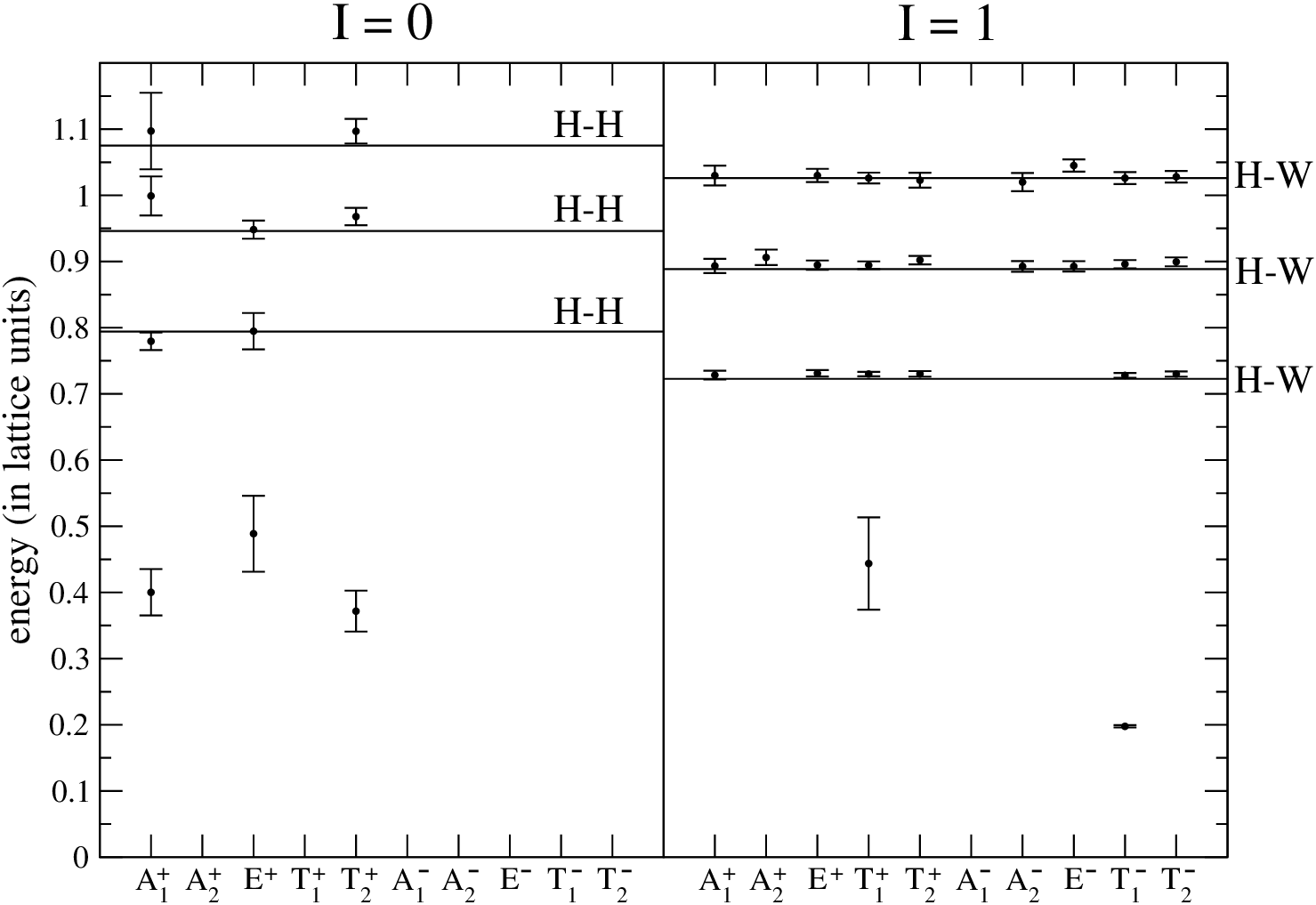}
\caption{Energy spectrum extracted from correlation functions of Higgs-Higgs and Higgs-$W$ operators built from Eqs.~(\ref{Hp}) and (\ref{Wp}) with $\left|\vec{p}\right|=2\pi/L$, $\left|\vec{p}\right|=\sqrt{2}(2\pi/L)$ and $\left|\vec{p}\right|=\sqrt{3}(2\pi/L)$ on a $24^3\times 48$ lattice with $\beta=8$, $\kappa=0.131$ and $\lambda=0.0033$.  Data points are lattice results with statistical errors; horizontal lines are the expectations from Eq.~\eqref{lattice_dispersion}.}
\label{graph_mass_hh_hw_momentum}
\end{figure}

\begin{figure}[tb]
\centering
\includegraphics[scale=0.6,clip=true]{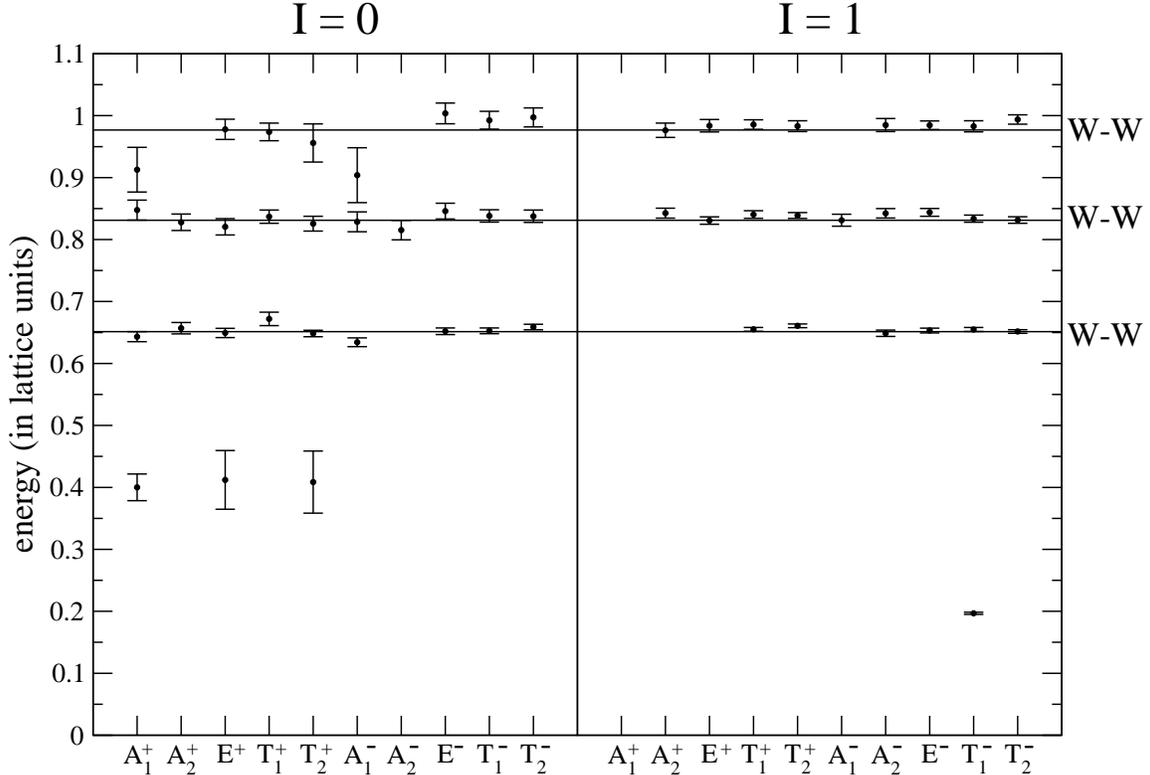}
\caption{Energy spectrum extracted from correlation functions of $W$-$W$ operators built from Eq.~(\ref{Wp}) with $\left|\vec{p}\right|=2\pi/L$, $\left|\vec{p}\right|=\sqrt{2}(2\pi/L)$ and $\left|\vec{p}\right|=\sqrt{3}(2\pi/L)$ on a $24^3\times 48$ lattice with $\beta=8$, $\kappa=0.131$ and $\lambda=0.0033$.  Data points are lattice results with statistical errors; horizontal lines are the expectations from Eq.~\eqref{lattice_dispersion}.}
\label{graph_mass_ww_momentum}
\end{figure}

\begin{table}[tb]
\caption{Octahedral group multiplicities of Higgs-Higgs, Higgs-$W$ and $W$-$W$ operators built of the operators in Eqs.~(\ref{Hp}) and (\ref{Wp}) with $\vec p\neq\vec 0$, where $\vec{p}_1=\tfrac{2\pi}{L}(1,0,0)$, $\vec{p}_2=\tfrac{2\pi}{L}(0,1,0)$ and $\vec{p}_3=\tfrac{2\pi}{L}(0,0,1)$.  Repeated SU(2) indices $a$, $b$, $c$ are summed, but Lorentz indices $\mu$, $\nu$, $\rho$ are not.  The indices $\mu$, $\nu$, $\rho$ are not equal to one another.}
\label{momentum1_multiplicities}
\begin{tabular}{r|c|ccccccccccc} 
Operator                                                   &$~I~$&$A_1^+$&$A_2^+$&$E^+$&$T_1^+$&$T_2^+$&$A_1^-$&$A_2^-$&$E^-$&$T_1^-$&$T_2^-$\\
\hline 
$H(\vec{p}_\mu)H(-\vec{p}_\mu)$                            &  0  &   1   &   0   &  1  &   0   &   0   &   0   &   0   &  0  &   0   &   0   \\
\hline
$H(\vec{p}_\mu)W_\mu^a(-\vec{p}_\mu)$                      &  1  &   1   &   0   &  1  &   0   &   0   &   0   &   0   &  0  &   1   &   0   \\
$H(\vec{p}_\mu)W_\nu^a(-\vec{p}_\mu)$                      &  1  &   0   &   0   &  0  &   1   &   1   &   0   &   0   &  0  &   1   &   1   \\
\hline
$W_\mu^a(\vec{p}_\mu)W_\mu^a(-\vec{p}_\mu)$                &  0  &   1   &   0   &  1  &   0   &   0   &   0   &   0   &  0  &   0   &   0   \\
$W_\nu^a(\vec{p}_\mu)W_\nu^a(-\vec{p}_\mu)$                &  0  &   1   &   1   &  2  &   0   &   0   &   0   &   0   &  0  &   0   &   0   \\
$W_\mu^a(\vec{p}_\mu)W_\nu^a(-\vec{p}_\mu)$                &  0  &   0   &   0   &  0  &   1   &   1   &   0   &   0   &  0  &   1   &   1   \\
$W_\nu^a(\vec{p}_\mu)W_\rho^a(-\vec{p}_\mu)$               &  0  &   0   &   0   &  0  &   0   &   1   &   1   &   0   &  1  &   0   &   0   \\
\hline
$\epsilon^{abc}W_\mu^b(\vec{p}_\mu)W_\mu^c(-\vec{p}_\mu)$  &  1  &   0   &   0   &  0  &   0   &   0   &   0   &   0   &  0  &   1   &   0   \\
$\epsilon^{abc}W_\nu^b(\vec{p}_\mu)W_\nu^c(-\vec{p}_\mu)$  &  1  &   0   &   0   &  0  &   0   &   0   &   0   &   0   &  0  &   1   &   1   \\
$\epsilon^{abc}W_\mu^b(\vec{p}_\mu)W_\nu^c(-\vec{p}_\mu)$  &  1  &   0   &   0   &  0  &   1   &   1   &   0   &   0   &  0  &   1   &   1   \\
$\epsilon^{abc}W_\nu^b(\vec{p}_\mu)W_\rho^c(-\vec{p}_\mu)$ &  1  &   0   &   0   &  0  &   1   &   0   &   0   &   1   &  1  &   0   &   0   \\
\end{tabular} 
\end{table}
\begin{table}[tb]
\caption{Octahedral group multiplicities of Higgs-Higgs, Higgs-$W$ and $W$-$W$ operators built of the operators in Eqs.~(\ref{Hp}) and (\ref{Wp}) with $\vec p\neq\vec 0$, where $\vec{p}_{12}=\tfrac{2\pi}{L}(1,1,0)$, $\vec{p}_{23}=\tfrac{2\pi}{L}(0,1,1)$, $\vec{p}_{31}=\tfrac{2\pi}{L}(1,0,1)$, $\vec{p}_{1-2}=\tfrac{2\pi}{L}(1,-1,0)$, $\vec{p}_{2-3}=\tfrac{2\pi}{L}(0,1,-1)$ and $\vec{p}_{3-1}=\tfrac{2\pi}{L}(-1,0,1)$.  Repeated SU(2) indices $a$, $b$, $c$ are summed, but Lorentz indices $\mu$, $\nu$, $\rho$ are not.  The indices $\mu$, $\nu$, $\rho$ are not equal to one another.}
\label{momentum2_multiplicities}
\begin{tabular}{r|c|ccccccccccc} 
Operator                                                               &$~I~$&$A_1^+$&$A_2^+$&$E^+$&$T_1^+$&$T_2^+$&$A_1^-$&$A_2^-$&$E^-$&$T_1^-$&$T_2^-$\\
\hline 
$H(\vec{p}_{\mu\nu})H(-\vec{p}_{\mu\nu})$                              &  0  &   1   &   0   &  1  &   0   &   1   &   0   &   0   &  0  &   0   &   0   \\
\hline
$H(\vec{p}_{\mu\nu})W_\mu^a(-\vec{p}_{\mu\nu})$                        &  1  &   1   &   1   &  2  &   1   &   1   &   0   &   0   &  0  &   2   &   2   \\
$H(\vec{p}_{\mu\nu})W_\rho^a(-\vec{p}_{\mu\nu})$                       &  1  &   0   &   0   &  0  &   1   &   1   &   0   &   1   &  1  &   1   &   0   \\
\hline
$W_\mu^a(\vec{p}_{\mu\nu})W_\mu^a(-\vec{p}_{\mu\nu})$                  &  0  &   1   &   1   &  2  &   1   &   1   &   0   &   0   &  0  &   0   &   0   \\
$W_\rho^a(\vec{p}_{\mu\nu})W_\rho^a(-\vec{p}_{\mu\nu})$                &  0  &   1   &   0   &  1  &   0   &   1   &   0   &   0   &  0  &   0   &   0   \\
$W_\mu^a(\vec{p}_{\mu\nu})W_\nu^a(-\vec{p}_{\mu\nu})$                  &  0  &   1   &   0   &  1  &   0   &   1   &   0   &   0   &  0  &   1   &   1   \\
$W_\mu^a(\vec{p}_{\mu\nu})W_\rho^a(-\vec{p}_{\mu\nu})$                 &  0  &   0   &   0   &  0  &   2   &   2   &   1   &   1   &  2  &   1   &   1   \\
\hline
$\epsilon^{abc}W_\mu^b(\vec{p}_{\mu\nu})W_\mu^c(-\vec{p}_{\mu\nu})$    &  1  &   0   &   0   &  0  &   0   &   0   &   0   &   0   &  0  &   1   &   1   \\
$\epsilon^{abc}W_\rho^b(\vec{p}_{\mu\nu})W_\rho^c(-\vec{p}_{\mu\nu})$  &  1  &   0   &   0   &  0  &   0   &   0   &   0   &   0   &  0  &   1   &   1   \\
$\epsilon^{abc}W_\mu^b(\vec{p}_{\mu\nu})W_\nu^c(-\vec{p}_{\mu\nu})$    &  1  &   0   &   1   &  1  &   1   &   0   &   0   &   0   &  0  &   1   &   1   \\
$\epsilon^{abc}W_\mu^b(\vec{p}_{\mu\nu})W_\rho^c(-\vec{p}_{\mu\nu})$   &  1  &   0   &   0   &  0  &   2   &   2   &   1   &   1   &  2  &   1   &   1   \\
\end{tabular} 
\end{table}
\begin{table}[tb]
\caption{Octahedral group multiplicities of Higgs-Higgs, Higgs-$W$ and $W$-$W$ operators built of the operators in Eqs.~(\ref{Hp}) and (\ref{Wp}) with $\vec p\neq\vec 0$, where $\vec{p}_{123}=\tfrac{2\pi}{L}(1,1,1)$, $\vec{p}_{-123}=\tfrac{2\pi}{L}(-1,1,1)$, $\vec{p}_{1-23}=\tfrac{2\pi}{L}(1,-1,1)$ and $\vec{p}_{12-3}=\tfrac{2\pi}{L}(1,1,-1)$.  Repeated SU(2) indices $a$, $b$, $c$ are summed, but Lorentz indices $\mu$, $\nu$, $\rho$ are not.  The indices $\mu$, $\nu$, $\rho$ are not equal to one another.}
\label{momentum3_multiplicities}
\begin{tabular}{r|c|ccccccccccc} 
Operator                                                                     &$~I~$&$A_1^+$&$A_2^+$&$E^+$&$T_1^+$&$T_2^+$&$A_1^-$&$A_2^-$&$E^-$&$T_1^-$&$T_2^-$\\
\hline 
$H(\vec{p}_{\mu\nu\rho})H(-\vec{p}_{\mu\nu\rho})$                            &  0  &   1   &   0   &  0  &   0   &   1   &   0   &   0   &  0  &   0   &   0   \\
\hline
$H(\vec{p}_{\mu\nu\rho})W_\mu^a(-\vec{p}_{\mu\nu\rho})$                      &  1  &   1   &   0   &  1  &   1   &   2   &   0   &   1   &  1  &   2   &   1   \\
\hline
$W_\mu^a(\vec{p}_{\mu\nu\rho})W_\mu^a(-\vec{p}_{\mu\nu\rho})$                &  0  &   1   &   0   &  1  &   1   &   2   &   0   &   0   &  0  &   0   &   0   \\
$W_\mu^a(\vec{p}_{\mu\nu\rho})W_\nu^a(-\vec{p}_{\mu\nu\rho})$                &  0  &   1   &   0   &  1  &   1   &   2   &   1   &   0   &  1  &   1   &   2   \\
\hline
$\epsilon^{abc}W_\mu^b(\vec{p}_{\mu\nu\rho})W_\mu^c(-\vec{p}_{\mu\nu\rho})$  &  1  &   0   &   0   &  0  &   0   &   0   &   0   &   1   &  1  &   2   &   1   \\
$\epsilon^{abc}W_\mu^b(\vec{p}_{\mu\nu\rho})W_\nu^c(-\vec{p}_{\mu\nu\rho})$  &  1  &   0   &   1   &  1  &   2   &   1   &   0   &   1   &  1  &   2   &   1   \\
\end{tabular} 
\end{table}
Tables~\ref{momentum1_multiplicities}, \ref{momentum2_multiplicities} and \ref{momentum3_multiplicities} show the multiplicities for Higgs-Higgs, Higgs-$W$ and $W$-$W$ operators with the nonzero internal momentum, $\left|\vec{p}\right|=2\pi/L$, $\left|\vec{p}\right|=\sqrt{2}(2\pi/L)$ and $\left|\vec{p}\right|=\sqrt{3}(2\pi/L)$, respectively.  The list of allowed $W$-$W$ representations for $\left|\vec{p}\right|=2\pi/L$ agrees completely with the states that were found in Figs.~\ref{graph_mass_link} and~\ref{graph_mass_wilson_polyakov}.
This shows why the $W$-$W$ signal was absent from other channels in those graphs.
In general, the direction of the internal momentum on the lattice will affect the allowed irreducible representations of multiparticle states \cite{Moore:2006ng,Moore:2005dw}.
Application of the variational analysis to the two-Higgs, Higgs-$W$ and two-$W$ operators with back-to-back momenta $\left|\vec{p}\right|=2\pi/L$, $\left|\vec{p}\right|=\sqrt{2}(2\pi/L)$ and $\left|\vec{p}\right|=\sqrt{3}(2\pi/L)$ produced Figs.~\ref{graph_mass_hh_hw_momentum} and \ref{graph_mass_ww_momentum}.

The single-$W$ states (near energy 0.2) and two-stationary-$W$ states (near 0.4) were detected in a few
channels but, as intended, these operators couple strongly to a pair with internal momentum.  Comparison of Tables~\ref{momentum1_multiplicities}, \ref{momentum2_multiplicities} and \ref{momentum3_multiplicities} with Figs.~\ref{graph_mass_hh_hw_momentum} and~\ref{graph_mass_ww_momentum} shows that signals are observed in precisely the expected subset of $I(\Lambda^P)$ channels in each case.

\clearpage

\section{Conclusions}\label{sec:conclusions}

The particle spectrum of the SU(2)-Higgs model has been computed thoroughly, using lattice simulations with all parameters tuned to experimental values.
Three conceptually different classes of operators were used to extract the energy spectrum: gauge-invariant links, Wilson loops and Polyakov loops.
Particular spatial shapes were chosen for these operators to provide access to all irreducible representations of angular momentum and parity, for both isospin 0 and 1.
Varying levels of stout-link and scalar smearing were applied to improve the operators and to generate a basis for a variational analysis of the correlation matrices.
The energies computed from the variational analysis comprise a vast multi-particle spectrum that is completely consistent with collections of almost-noninteracting Higgs and $W$ bosons.  No states were found beyond this simple picture.

Of course the interactions between bosons are not expected to be strictly zero,
but such tiny deviations from zero are not attainable using the lattice studies presented here.
Simulations with a stronger gauge coupling -- but still in the Higgs region
of the phase diagram -- might provide information about
interactions, and the fact that the SU(2)-Higgs model is a single phase
implies an analytic connection from strong coupling to the physical point.
It also implies an analytic connection to the confinement region of the phase
diagram with its seemingly very different spectrum.
Therefore future lattice studies, similar to what we have done but
at stronger gauge coupling, could be of significant value.

Our study, by observing more than a dozen distinct energy levels from the single $W$ up to multiboson states with various momentum options, represents a major step beyond previous simulations of this spectrum.
Our work demonstrates that present-day lattice methods can provide
precise quantitative results for the Higgs-$W$ boson spectrum.

\section*{Acknowledgments}

The authors thank Colin Morningstar for helpful discussions about the smearing of
lattice operators.
This work was supported in part by the Natural Sciences and
Engineering Research Council (NSERC) of Canada, and by computing resources of
WestGrid\cite{westgrid} and SHARCNET\cite{sharcnet}.

\end{document}